\documentclass[pra,twocolumn,showpacs,amsmath,amssymb,amsfonts,nofootinbib]{revtex4}

\newtheorem{theorem}{Theorem}

\newtheorem{assumption}{Assumption}
\newtheorem{corollary}{Corollary}
\newtheorem{definition}{Definition}
\newtheorem{constraint}{Constraint}
\newtheorem{conjecture}{Conjecture}

\usepackage{graphicx}
\usepackage{psfrag}

\begin{document}

\title{Information processing in generalized probabilistic theories}

\author{Jonathan Barrett}
\email{jbarrett@perimeterinstitute.ca}
\affiliation{Perimeter Institute for Theoretical Physics, 31 Caroline Street N, Waterloo, Ontario N2L 2Y5, Canada}

\begin{abstract}
I introduce a framework in which a variety of probabilistic theories can be defined, including classical and quantum theories, and many others. From two simple assumptions, a tensor product rule for combining separate systems can be derived. Certain features, usually thought of as specifically quantum, turn out to be generic in this framework, meaning that they are present in all except classical theories. These include the non-unique decomposition of a mixed state into pure states, a theorem involving disturbance of a system on measurement (suggesting that the possibility of secure key distribution is generic), and a no-cloning theorem. Two particular theories are then investigated in detail, for the sake of comparison with the classical and quantum cases. One of these includes states that can give rise to arbitrary non-signalling correlations, including the super-quantum correlations that have become known in the literature as Nonlocal Machines or Popescu-Rohrlich boxes. By investigating these correlations in the context of a theory with well-defined dynamics, I hope to make further progress with a question raised by Popescu and Rohrlich, which is, why does quantum theory not allow these strongly nonlocal correlations? The existence of such correlations forces much of the dynamics in this theory to be, in a certain sense, classical, with consequences for teleportation, cryptography and computation. I also investigate another theory in which all states are local. Finally, I raise the question of what further axiom(s) could be added to the framework in order uniquely to identify quantum theory, and hypothesize that quantum theory is optimal for computation.  
\end{abstract}
\maketitle

\section{Introduction}\label{sec:introduction}

The question is periodically raised, what is responsible for the power of quantum computation (or cryptography, or information processing in general)? At a recent meeting in Konstanz \cite{konstanz}, speakers referred to quantum entanglement; the superposition principle; the exponentially growing size of Hilbert space with the number of qubits; nonlocality and contextuality; the possibility of continuous reversible transformations between pure states; and the so-called sign problem in Monte Carlo simulations of certain types of quantum system \cite{troyer05}. It is perhaps unsurprising that there are so many different answers. The problem is that the results of quantum information theory are already well understood as consequences of the quantum formalism, and it is not clear that simply pointing to aspects of that formalism tells us anything new.
What we are really looking for is a better understanding of the connections between information processing and physical principles in general. 

Such an understanding could be gained by studying information processing in a broader range of theories than classical and quantum, where different physical principles may hold. For any theory, whether it applies to Nature or not, one can consider the information processing possibilities of this theory, the differences from those of classical or quantum theory, and attempt to trace these possibilities back to the fundamental features of the theory. Some authors have indeed investigated unrealistic theories, with a view to understanding the relevant features \cite{abrams98, aaronson, aaronson1,aaronson3, brun, bacon, aaronsoninprep, hardyteleport, spekkens, smolin, masanes}.

To make further progress along these lines, I introduce an operational framework for probabilistic theories in which a broad range of different theories can be defined. The framework, described in Sections~\ref{sec:fram-prob-theor}, \ref{ambiguities} and \ref{sec:some-diff-theor}, is based on that used by Hardy in his derivation of quantum theory from simple axioms \cite{hardy}. The basic idea is that a state is represented as a vector of probabilities of measurement outcomes. Transformations of a system must correspond to linear transformations of this vector. By including probabilistic, that is normalization-decreasing, transformations, a unified account of transformations and measurements can be given. Rather than employ any of Hardy's axioms, I introduce two assumptions that concern how separate systems combine to form a joint system. The first is that operations on the separate systems commute (this implies a no-signalling principle), and the second is that the state of the joint system can be completely specified by joint probabilities for local measurements. From these assumptions a tensor product rule can be derived. This removes at least some of the mystery from the quantum tensor product rule and generalizes a derivation by Fuchs \cite{fuchs}.

The resulting framework includes classical probabilistic theories, quantum theory, and many other theories besides. The first thing one notices is that certain phenomena, usually thought of as specifically quantum, are in fact generic. This means that they either appear in all theories, or they appear in all theories except classical theories, which emerge as a very special case. As shown in Section~\ref{sec:some-basic-prop}, these phenomena include the non-unique decomposition of a mixed state into pure states, a theorem concerning the disturbance of a system on measurement, and the no-cloning theorem. (These observations are complementary to those of Ref.~\cite{masanes}, where it is noted that similar properties hold in nonlocal but non-signalling theories.) 

In addition to looking at generic properties of theories, it is useful to analyze at least one or two novel theories in detail. These then provide well-understood examples that can be contrasted with the classical and quantum cases. Thus the rest of this work is devoted to an analysis of two theories that admit a particularly natural definition. The first of these allows arbitrary correlations between measurements on separated systems, as long as they are non-signalling. I call it Generalized Non-Signalling Theory (GNST). The correlations allowed by this theory can be more nonlocal than quantum theory allows, and include the super-quantum correlations that have come to be known variously in the literature as Popescu-Rohrlich (PR) boxes, or Nonlocal Machines \cite{kt,pr,barrettetal,boxot,short,boxcommit,broadbentmethot,barrettpironio,jonesmasanes,vandam,brassardetal}. Popescu and Rohrlich raised the question of why quantum theory does not allow these correlations. An investigation of a complete theory, with dynamics, that does include the correlations may help to answer this question. The second theory allows the same states of single systems as GNST, but does not allow any violation of Bell inequalities. For this reason it is called Generalized Local Theory (GLT).

One of the interesting things about GNST is that there are many direct analogues of quantum phenomena (in addition to the generic phenomena mentioned above). These include entanglement, nonlocality, a form of contextuality, and the Einstein-Podolski-Rosen (EPR) paradox. (Interestingly, a quite different toy theory introduced by Spekkens displays many of these phenomena too \cite{spekkens}.) However, there are also differences with quantum theory. A central insight of this work is that there is a trade-off between the allowed states of a theory and the allowed dynamics. This follows from the simple fact that dynamics has to act in such a way that allowed states are taken to allowed states. In the case of GNST, the fact that all non-signalling correlations are possible means that the dynamics is highly restricted. In fact, I show in Section~\ref{dynamics} that the dynamics of single systems in GNST is essentially classical, corresponding to no more than relabellings of measurements and outcomes. This result is extended to transformations and measurements on simple kinds of bipartite systems (more complicated cases are still open). GLT is in some sense intermediate, with transformations on single systems similarly simple, but with transformations on bipartite systems including other possibilities.

These conclusions about dynamics have consequences for information processing, discussed in Section~\ref{informationprocessing}. For example, there is no teleportation in GNST, despite the existence of highly nonlocal states that might have been thought to facilitate a task like teleportation. Key distribution is possible in GNST and 1-2 oblivious transfer in both GNST and GLT. Other cryptographic possibilities, such as key distribution in GLT, or bit commitment in either theory, are open questions. A natural circuit-type model of computation can be defined for any theory in the framework. The states and dynamics together in GLT are sufficiently restricted that computation can be simulated efficiently by a classical computer. The theorems concerning dynamics in GNST give evidence that computation in this theory can also be simulated efficiently by a classical computer (\emph{despite} the existence of super-entangled states). The fact that quantum theory, unlike GNST and GLT, achieves such a harmonious balance of states and dynamics leads to the following hypothesis that I leave open: \emph{a quantum computer can simulate computation in any theory in the framework with at most polynomial overhead}.

Finally, two motivations are not directly connected with information processing. On the face of it, most of the theories that can be written down in the framework described suffer from similar interpretational problems as quantum theory. For example, are pure states in one of these theories best regarded as complete descriptions of individual reality, as describing only ensembles, or as descriptions of agents' degrees of belief? Although I do not do this in this paper, consideration of these questions in a broader framework may shed new light on the quantum theoretical problems. The other motivation is to stimulate research into finding ways of deriving quantum theory from physical principles (instead of laying down a list of mathematical axioms, as per the standard textbook approach). What principles could be used to rule out the other theories described and leave only quantum theory? One reason for deriving quantum theory from physical principles is that by modifying one or another of the principles, we may discover new ways of going beyond quantum theory.

\section{A framework for probabilistic theories}\label{sec:fram-prob-theor}

This section describes in some detail a general operational framework in which probabilistic theories can be written down. All theories in this framework share the following features with classical and quantum theory.
\begin{enumerate}
\item Local operations on distinct subsystems commute. In the case of a bipartite system $AB$, for example, this means that if an operation is performed on system $A$ alone, and an operation on system $B$ alone, it does not matter what order the operations were performed in.
\item The global state of a composite system is determined by correlations between local measurements.
\end{enumerate}

\subsection{States and Operations}\label{sec:stat-transf}

Consider a laboratory containing \emph{preparation devices} and \emph{operation devices}. Preparation devices prepare a system in a given state and operation devices act on a system, in general changing its state. When an operation device is used, there may be several different outcomes, each occurring with some probability. Each outcome is identified by a different macroscopic event (for example, a different light being illuminated on the device, or a different position of a pointer). Thus operation devices serve to perform both transformations and measurements. Given the state of a system, it should be possible to calculate the probabilities of measurement outcomes for any measurement. Conversely, if the probabilities of measurement outcomes for any measurement are known, then the state is known.   

Suppose that systems come in different \emph{types}, where in quantum theory, for example, the type of system corresponds to the dimension of its Hilbert space. For each type of system, there is some finite set ${\cal F}$ of measurements, each with a finite number of outcomes, such that the state of the system can be completely specified by listing the probabilities for these outcomes. For example, in quantum theory, the state of a spin-$1/2$ particle can be specified by giving the probabilities of obtaining spin-up on measuring in the $x$, $y$ and $z$ directions. Call the measurements in ${\cal F}$ \emph{fiducial} measurements and ${\cal F}$ the \emph{fiducial set}. In general, there will be other measurements that can be performed on a system that are not contained in the fiducial set (a measurement of spin in some direction at $45^{\circ}$ to the $z$-axis, say). The probabilities of outcomes of these measurements can nevertheless be determined from the state. We ignore the possibility of states requiring an infinite number of probabilities to be specified (despite the fact that quantum theory includes infinite dimensional systems and classical probability theory infinite sample sets). The set of fiducial measurements need not be unique. In general it will be possible to find a different set (perhaps involving a different number of measurements with different numbers of outcomes) that also suffices to specify the state. 

This is essentially the framework described by Hardy \cite{hardy}, who introduced the term \emph{fiducial} for the state-defining measurements. (See also \cite{wootters,fuchs,mana1,mana2}, where the idea of representing a state via probabilities for measurement outcomes is also explored.) Unlike Hardy, we shall assume for convenience that the degrees of freedom expressed in the state are internal degrees of freedom, and that all measurements are measurements of internal degrees of freedom. With respect to spacetime degrees of freedom, systems behave classically, having a definite position and velocity at all times. This seems the most natural position to take given that we are most interested in the information processing properties of the different theories considered. However, it would be interesting to extend this work, and to consider what Nature would be like if all degrees of freedom, including those of spacetime, were described by a theory like one of the ones presented here (but extended to allow for infinite-outcome measurements).  

The above is summarized by
\begin{assumption}
The state of a single system can be completely specified by listing the probabilities for the outcomes of some subset ${\cal F}$ of all possible measurements. These are the fiducial measurements. These probabilities can be written arranged in a vector.
\begin{equation}
\vec{P} \equiv \left(\begin{array}{c} P(a=1|X=1)\\P(a=2|X=1)\\ \vdots \\ \hline P(a=1|X=2) \\ P(a=2|X=2) \\ \vdots \\ \hline \vdots \end{array}\right).
\end{equation}  
$P(a=i|X=j)$ is the probability of getting outcome $i$ when fiducial measurement $j\in {\cal F}$ is performed on the system. 
\end{assumption}
Normalization of the state would require that
\begin{equation}\label{absolutenorm}
\sum_i P(a=i|X=j)=1 \qquad \forall j,
\end{equation}
where the sum ranges over all the values $i$ that the outcome can take for a particular measurement. It is convenient also to give a meaning to unnormalized states (just as in quantum theory it is sometimes convenient to write down unnormalized density matrices). Suppose that a system is prepared in some (normalized) state and an operation performed with an outcome $i$ that is obtained with probability less than $1$. There is an unnormalized state associated with $i$, each entry of which is the joint probability of getting $i$ followed by a particular outcome for a subsequent fiducial measurement. This implies that unnormalized states satisfy
\begin{equation}\label{relativenorm}
\sum_{i'} P(a=i'|X=j)=\sum_{i''} P(a=i''|X=j')=c \quad \forall j,j'
\end{equation}
with $0\leq c \leq 1$. In the case described, $c$ is the probability of the outcome $i$. This idea generalizes to chains of operations, thus operations should be defined on unnormalized states as well as on normalized ones. Define
\begin{equation}
|\vec{P}|\equiv \sum_i P(a=i|X=j),
\end{equation}
where the right hand side is independent of the choice of $j$. The notation $|\vec{P}|$ is used throughout and should not be confused with more usual definitions of the norm of a vector.

Suppose that for each type of single system, the fiducial measurements are fixed. A particular theory will specify, for each type of system, a set of allowed vectors $\vec{P}$. These correspond to physically possible states of a system, i.e., states that can actually be prepared using one of the preparation devices. There is no reason to suppose that all vectors $\vec{P}$ that can be written down can actually be prepared. For example, in quantum theory, one cannot prepare a system that will with certainty return the outcome spin-up for spin measurements in both the $z$- and $x$-directions. Call the set of allowed states ${\cal S}$ (where there is a different ${\cal S}$ for each type of system but we suppress this dependence).
\begin{assumption}\label{convexsassumption}
For each type of system, the set of allowed normalized states is closed and convex. The complete set of states ${\cal S}$ is the convex hull of the set of allowed normalized states and $\vec{0}$.
\end{assumption}
$\vec{0}$ is the vector with all entries $0$. The idea behind this assumption is that it is always possible to toss a biased coin and subsequently to be interested only in the joint probabilities of getting given measurement outcomes along with heads. In this way one can `prepare' unnormalized states. If heads occurs with probability zero, the state $\vec{0}$ is prepared. Convexity of ${\cal S}$ corresponds to the assumption that if it is possible to prepare states $P_1$ and $P_2$, then it is also possible to prepare any probabilistic mixture of the two states. One may toss a coin, prepare either $P_1$ or $P_2$ depending on the outcome, and then forget the outcome.\footnote{The assumption is also stated in such a manner as to rule out the possibility of an unnormalized state without a corresponding normalized state.} Extreme points of $S$ apart from $\vec{0}$ are \emph{pure states}. States that are neither pure nor $\vec{0}$ are \emph{mixed}. Mixed states can be written as a convex sum of pure states and $\vec{0}$, but this sum need not be unique.

Notice from Eq.~\eqref{relativenorm} that ${\cal S}$ lies in a subspace of the complete vector space. In general, we allow for the possibility that $\vec{P}$ is an over-complete description of the state of a system. Thus there may be other linear constraints that apply apart from Eq.~\eqref{relativenorm} implying that ${\cal S}$ lies in a smaller subspace still.

When an operation is performed, each outcome is associated with a transformation of the state of the system, i.e., with a map from states to states:
\begin{equation}
\vec{P}\rightarrow \vec{P}'=f(\vec{P}).
\end{equation}
Some operations have only one outcome and the corresponding transformation preserves normalization of the state (in quantum theory, these are the trace-preserving completely positive maps).  If an outcome occurs with probability $<1$, then it is associated with a transformation that decreases the normalization of the state (in quantum theory, these are trace-decreasing completely positive maps). In the most general case, one could consider operations that change the system into a system of a different type (just as in quantum theory one sometimes considers completely positive maps between Hilbert spaces of different dimension). In this work I assume that operations do not change the type of system, although the appropriate generalization is not usually too difficult.  
 
Consider a transformation acting on a system that is in a mixed state, that is a state $\vec{P}$ such that
\begin{equation}
\vec{P} = \sum_i q_i \vec{P}_i,
\end{equation}
where the $\vec{P}_i$ are allowed states and where $0\leq q_i \leq 1$ and $\sum_i q_i = 1$. One way of preparing a system in such a state would be to prepare a system in the state $\vec{P}_i$ with probability $q_i$ and then to forget the value of $i$. In this case the transformed $\vec{P}$ must be the same convex combination of the transformed $\vec{P}_i$, that is
\begin{equation}\label{preservecombinations}
f(\vec{P}) = f\left(\sum_i q_i \vec{P}_i\right) = \sum_i q_i f(\vec{P_i}) \qquad \forall P_i\in{\cal S}.
\end{equation}
It follows from this that the action of $f$ on the set of allowed states $\vec{P}$ can be represented as 
\begin{equation}
\vec{P} \rightarrow M.\vec{P},
\end{equation}
where $M$ is a matrix, i.e, $f$ is a linear map. This is not completely obvious from Eq.(\ref{preservecombinations}), since the equation involves only convex combinations, and furthermore only applies for those $\vec{P}_i \in {\cal S}$. A rigorous proof is given in Appendix~\ref{linearityproof}. 

An operation corresponds to a set of matrices $\{M_i\}$.\footnote{A note on terminology. I shall continue to use the term \emph{operation} to refer to the experiment with a number of different outcomes corresponding to the set $\{M_i\}$, and the term \emph{transformation} to refer to a single, in general normalization-decreasing, $M_i$.} The unnormalized state associated with the $i$th outcome is $M_i.\vec{P}$, and the probability of the $i$th outcome is
\begin{equation}
\frac{|M_i.\vec{P}|}{|\vec{P}|}.
\end{equation}
For each type of system, a particular theory will specify a set of allowed operations. Denote this set ${\cal O}$. An element of ${\cal O}$ is a set of transformations $\{M_i\}$, and must be such that the following holds.
\begin{constraint}\begin{align}
0\leq \frac{|M_i.\vec{P}|}{|\vec{P}|} \leq 1 \qquad  &\forall i,\vec{P}\in {\cal S},\label{allowedopconditions1}\\
\sum_i \frac{|M_i.\vec{P}|}{|\vec{P}|} = 1 \qquad &\forall \vec{P}\in {\cal S},\label{allowedopconditions2}\\
M_i.\vec{P} \in {\cal S}\qquad  &\forall i,\vec{P}\in{\cal S}.
\end{align}\end{constraint}
A further constraint is that each transformation $M_i$ must result only in allowed states when it acts on a system that is part of a larger multi-partite system (see next section). The following assumption results in some loss of generality but also makes things simpler.
\begin{assumption}
For each type of system, there is a set ${\cal T}$ of allowed transformations. A set of transformations $\{M_i\}$ is an element of ${\cal O}$ if and only if $M_i\in {\cal T}\ \forall i$, and Eq.~\eqref{allowedopconditions2} is satisfied. The set ${\cal T}$ includes the transformation that maps all $\vec{P}$ to $\vec{0}$ and is convex.
\end{assumption}
With this assumption, once ${\cal T}$ is given, a separate specification of ${\cal O}$ is not needed. The reasons for convexity are similar to those given for Assumption~\ref{convexsassumption}.

As mentioned above, the formalism of operations already includes measurements. Sometimes one is not interested in the state after measurement but only in the probabilities of the different outcomes. In this case it is convenient to associate with an operation $\{M_i\}$ a set of vectors $\{R_i\}$ such that
\begin{equation}\label{ridefinition}
\vec{R_i}.\vec{P}= |M_i.\vec{P}|\ \forall \vec{P}\in{\cal S}.
\end{equation}
Such a set can always be found. For a normalized $\vec{P}$, the probability of the $i$th outcome is then given by $\vec{R}_i.\vec{P}$. It does not matter if the vector $\vec{R}_i$ is not unique - this simply means that different vectors can represent the same measurement outcome. Denote by ${\cal M}$ the set of all sets $\{R_i\}$ such that Eq.~\eqref{ridefinition} holds for some $\{M_i\}\in {\cal O}$. ${\cal M}$ is the set of allowed measurements. Denote by ${\cal R}$ the set of allowed measurement vectors, that is, the set of vectors $\vec{R}$ such that $\vec{R}.\vec{P}= |M.\vec{P}|\ \forall \vec{P}\in{\cal S}$, for some $M\in {\cal T}$.\footnote{Recall that in quantum theory, an effect $E$ is a positive operator such that $0\leq E\leq 1$. $\vec{R}$ vectors are essentially a generalization of the effects to our framework. In the usual quantum formalism, an effect can represent a yes/no measurement on a quantum state $\rho$, with the probability of the yes outcome given by $\mathrm{Tr}(E\rho)$. A set of effects $E_i$ such that $\sum_i E_i=I$, where $I$ is the identity, is a positive operator-valued (POV) decomposition of the identity, and corresponds to a POV measurement.} (Notation: ${\cal R}$ should not be confused with $\mathbb{R}$, the set of real numbers.)

\subsection{Multi-partite systems}\label{sec:multi-part-syst}

So far, the framework described is similar to that used by Hardy as a starting point for his derivation of quantum theory (although I have been more explicit about treating transformations and measurements in a unified manner). Hardy narrows things down with various axioms. Rather than adopt any of Hardy's axioms, however, I introduce a small number of non-trivial assumptions that concern how systems combine to make multi-partite systems. One reason for this is that most questions of information processing do not make sense without some notion of systems being composed of separate subsystems. From these assumptions I derive that systems combine according to a \emph{tensor product rule}. This is of independent interest since it sheds light on where this rule comes from in quantum theory.

From hereon, the notion of a \emph{type} of system is broadened. Thus multi-partite systems can come in different types, where a particular type of multi-partite system is composed of $n_A$ single systems of type $A$, $n_B$ single systems of type $B$, and so on. In all of this section, a system $A$ or $B$ refers to a system of some specific type, that may itself be a composite system.

Begin with the idea that, given a system $A$, it is possible to identify some operations as \emph{operations on system $A$ alone} and that, in particular, the fiducial measurements for system $A$ are operations on system $A$ alone. (Without this, one might say that we have no business speaking of separate systems in the first place.)
\begin{assumption}\label{localopscommute}
Local Operations Commute. Consider a joint system composed of systems $A$ and $B$. Suppose that an operation is performed on system $A$ alone with outcome $o_A$ and an operation on system $B$ alone with outcome $o_B$. The final unnormalized state of the joint system does not depend on the order in which the operations were performed. In particular, this implies that the joint probability of getting outcomes $o_A$ and $o_B$ does not depend on the ordering of the operations.
\end{assumption}
This assumption means that operations can be regarded as performed simultaneously on systems $A$ and $B$ without ambiguity. It also implies
\begin{corollary}\label{nosigprinciple}
The No-Signalling Principle. If an operation is performed on system $A$, it is not possible to get information about which operation was performed by measuring system $B$.
\end{corollary}
The proof of the corollary is straightforward. Suppose that an operation is performed on system $A$ first, followed by an operation on system $B$. Whichever operation was performed on system $A$, the marginal probability of outcome $o_B$ is equal to the probability of $o_B$ in the case that the operation on system $B$ came first. The probability of $o_B$ is thus independent of the operation on system $A$.  

\begin{assumption}\label{globalstateassumption}
The Global State Assumption. The global state of a multi-partite system can be completely determined by specifying joint probabilities of outcomes for fiducial measurements performed simultaneously on each subsystem.
\end{assumption}
Note that while the global state assumption is satisfied in quantum theory and in classical probability theory, it need not be satisfied in an arbitrary theory. For example, it is not true in the case of quantum theory defined over a real Hilbert space \cite{realhilbertspace,woottersreal,cavesfuchs}. So this assumption has significant content.\footnote{Arguably, this is not the case for the assumption that local operations commute, which may be regarded as part of the definition of what we mean by an operation being on system $A$ alone. Not wishing to be dogmatic on this point, I have listed this principle with the other assumptions. We should distinguish, however, the implied no-signalling principle from \emph{the impossibility of super-luminal signalling}, which is a contingent fact that as far as we know is true in our universe. To see the difference, consider that in the non-relativistic quantum mechanics of particles, the no-signalling principle is valid, yet super-luminal signalling is possible. In the present framework, the impossibility of superluminal signalling would imply an upper bound on the velocity of systems and that Alice cannot carry out an operation on Bob's system if she is spacelike separated from it. But I shall not use such notions, or indeed any notion of spacetime structure.} 

It follows from these two assumptions that the global state of a multi-partite system can be written in the form of a vector of joint probabilities. For example, for a bipartite system $AB$, it will look like this:
\begin{equation}
\vec{P}^{AB} \equiv \left(\begin{array}{c} P(a=1,b=1|X=1,Y=1)\\P(a=1,b=2|X=1,Y=1)\\ \vdots \\ \hline P(a=1,b=1|X=1,Y=2) \\ P(a=1,b=2|X=1,Y=2) \\ \vdots \\ \hline \vdots \end{array}\right).
\end{equation}  
$P(a=i,b=j|X=k,Y=l)$ is the joint probability of getting outcomes $i$ and $j$ when fiducial measurements $k$ and $l$ are performed on the two subsystems. The no-signalling principle implies
\begin{align}
&\begin{split} \sum_j  &P(a=i,b=j|X=k,Y=l) = \\
&\sum_{j'}  P(a=i,b=j'|X=k,Y=l')\quad \forall i,k,l,l',\end{split}\label{nosignalling1}\\
&\begin{split} \sum_i  &P(a=i,b=j|X=k,Y=l) = \\
&\sum_{i'}  P(a=i',b=j|X=k',Y=l)\quad \forall j,k,k',l.\end{split}\label{nosignalling2}
\end{align}

The \emph{reduced state} for system $A$ (analogous to the reduced state in quantum theory, or marginal probabilities in classical probability theory) is given by
\begin{equation}
\vec{P}^A = \left(\begin{array}{c} P(a=1|X=1)\\P(a=2|X=1)\\ \vdots \\ \hline P(a=1|X=2) \\ P(a=2|X=2) \\ \vdots \\ \hline \vdots \end{array}\right),
\end{equation}
where
\begin{equation}
P(a=i|X=j)=\sum_{i'} P(a=i,b=i'|X=j,Y=j').
\end{equation}
Here, $a$ and $X$ are the outcome and fiducial measurement for the system whose reduced state is defined, and $b$ and $Y$ are the outcome and fiducial measurement for the other system. The no-signalling conditions of Eqs.~\eqref{nosignalling1}, \eqref{nosignalling2} ensure that the sum on the right is independent of the choice of $j'$. 

As seen in the last section, a particular theory specifies a set ${\cal S}$ of allowed states for each type of system. This applies also for each type of multi-partite system. There is, however, a constraint. 
\begin{constraint}\label{reducedstateconstraint}
Suppose that $\vec{P}^{AB}\in {\cal S}^{AB}$, where ${\cal S}^{AB}$ is the set of allowed states for the joint system. Suppose that $\vec{P}^A$ is the reduced state for system $A$ corresponding to $\vec{P}^{AB}$. Then $\vec{P}^A\in {\cal S}^A$, where ${\cal S}^A$ is the set of allowed states for system $A$.
\end{constraint}

That systems combine according to a tensor product rule is asserted by the following three theorems. Proofs are in Appendix~\ref{tensorproductproof}.
\begin{theorem}\label{tensorproducttrivialtheorem}
Denote the vector spaces containing the vectors $\vec{P}^{AB}$, $\vec{P}^A$, and $\vec{P}^B$ by $V^{AB}$, $V^A$, and $V^B$ respectively. Then one can identify 
\begin{equation*}
V^{AB}=V^A\otimes V^B.
\end{equation*}
\end{theorem}
\begin{theorem}\label{directproducttheorem}
Any $\vec{P}^{AB}\in {\cal S}^{AB}$ can be written
\begin{equation}\label{directproduct}
\vec{P}^{AB} = \sum_i r_i \vec{P}_i^A\otimes \vec{P}_i^B,
\end{equation}
with the $r_i$ real, $\vec{P}^A_i\in {\cal S}^A$ and $\vec{P}^B_i\in {\cal S}^B$. Both $\vec{P}^A_i$ and $\vec{P}^B_i$ can be taken to be normalized and pure.
\end{theorem}
\begin{theorem}\label{tensorprodfortransforms}
Consider a transformation on system $A$ alone defined by
\begin{equation*}
\vec{P}^A\rightarrow \vec{P}^{\prime A} = M^A . \vec{P}^A.
\end{equation*}
The transformation of the joint system is given by
\begin{equation*}
\vec{P}^{AB}\rightarrow \vec{P}^{\prime AB} = (M^A\otimes I). \vec{P}^{AB}.
\end{equation*}
\end{theorem}
Recall that transformations include probabilistic transformations that decrease the normalization of the state. Thus an immediate corollary of Theorem~\ref{tensorprodfortransforms} is
\begin{corollary}
If a measurement is performed on system $A$ alone, with state $\vec{P}^A$, the probability of a particular outcome is given by
\begin{equation}
\vec{R}.\vec{P}^A = (\vec{R}\otimes\vec{I}).\vec{P}^{AB}.
\end{equation}
Here, $\vec{I}$ is a vector representing the identity measurement, that is $\vec{I}.\vec{P}^B=|\vec{P}^B| \ \forall \vec{P}^B\in{\cal S}^B$. The way things are set up, $\vec{I}$ is not unique but can always be taken to be $(1,\ldots,1|0,\ldots,0|0,\ldots,0|\cdots)$.
\end{corollary}

Much follows from these theorems and corollary. 

\emph{Collapsed states}. Suppose that an operation is performed on a system $A$ in a state $\vec{P}^A$. Suppose that the operation has outcomes $i$ such that the final normalized state conditioned on outcome $i$ is given by $\vec{P}^{A}_i\equiv M_i\vec{P}^A/|M_i\vec{P}^A|$. The change in the state of system $A$ is analogous to the quantum mechanical collapse of the state vector. If systems $A$ and $B$ begin in some joint state $\vec{P}^{AB}$, and a measurement is performed on system $A$, then the final state of system $B$, conditioned on a particular outcome for the measurement, is also unambiguously determined. Thus this ``collapse'' is also well-defined ``at a distance''. Typically, similar questions of interpretation arise in theories in this framework as do in quantum theory. Is this collapse a real process? A change in an agent's degrees of belief following her measurement? And so on.

\emph{Entanglement and nonlocality}. In Theorem~\ref{directproducttheorem}, a joint state of a system $AB$ is written as a linear sum of direct product states. Note that the theorem does not assert that a joint state of $AB$ can be written as a convex combination of direct product states. In general, there will be joint states that cannot be written in this form. These are the \emph{entangled} states of the theory. Entanglement is distinct from nonlocality, where the latter means violation of a Bell inequality. Thus i) there are theories such as classical theories that have no entanglement or nonlocality, ii) there may be theories that have entanglement but no nonlocality, and iii) there are theories, such as quantum theory and GNST developed below, that have both entanglement and nonlocality, although these may not coincide.\footnote{It is clear that entanglement is necessary for nonlocality. But in quantum theory there are entangled mixed states that are local \cite{werner,barrettmodel}, hence entanglement is not sufficient for nonlocality. In GNST, on the other hand, entanglement and nonlocality do coincide. This is because if one can write down a local model for a particular state in GNST, then the model will itself define a convex decomposition of that state into product states allowed by the theory. This is not true in quantum theory because arbitrary local models can employ probability assignments not corresponding to any quantum state.}

\emph{Multi-partite systems}. The state of a multi-partite system can be written as a vector $\vec{P}^{AB\ldots Z}\in V_A\otimes V_B \otimes\cdots\otimes V_Z$. This vector can be written as a linear sum of direct product states $\sum_i r_i \vec{P}^A_i\otimes \vec{P}^B_i\otimes\cdots\otimes\vec{P}^Z_i$, with $r_i\in\mathbb{R}$, $\vec{P}^A\in {\cal S}^A$, and so on. A transformation on system $A$ alone takes the form $M\otimes I\otimes\cdots\otimes I$, and similarly for transformations on $B,\ldots,Z$ alone. These extensions of the above theorems follow, since those theorems were stated for arbitrary bipartite systems $AB$ and included the fact that $A$ and $B$ may themselves be composite.  

Finally, recall that a theory, in addition to specifying the set ${\cal S}$ of allowed states for each type of system, must also specify the set ${\cal T}$ of allowed transformations. 
\begin{definition}\label{welldefined}
A transformation on system $A$ is \emph{well-defined} if $(M^A_i\otimes I).\vec{P}^{AB}\in {\cal S}^{AB}$ whenever $\vec{P}^{AB}\in{\cal S}^{AB}$, for all types of system $B$.
\end{definition}
This definition corresponds to the fact that in quantum theory, allowed transformations must be completely positive maps (and not, e.g., merely positive maps). An obvious constraint is 
\begin{constraint}\label{welldefinedconstraint}
For each type of system, all transformations $\in {\cal T}$ must be well-defined.
\end{constraint}
A natural assumption is
\begin{assumption}\label{directproductassumption}
If $\vec{P}^A\in {\cal S}^A$ and $\vec{P}^B\in {\cal S}^B$, then $\vec{P}^{A}\otimes \vec{P}^B \in {\cal S}^{AB}$.
\end{assumption}
A final assumption that is convenient is
\begin{assumption}\label{alltransformations}
A theory first specifies a set ${\cal S}$ of allowed states for each type of system. All transformations that are well-defined are then allowed transformations.
\end{assumption}
This assumption is indeed satisfied by all the theories considered below, including classical theories, quantum theory, GNST, and GLT. It is nice because it means that a theory is completely specified once the allowed types of system are specified, along with the set ${\cal S}$ of allowed states for each type. In this case, Assumption~\ref{alltransformations} defines the set ${\cal T}$. The way things are set up, each of the sets ${\cal O}$, ${\cal M}$ and ${\cal R}$ is in turn defined by ${\cal T}$. Assumption~\ref{alltransformations} also ensures that certain other obvious constraints hold that do not then need to be stated separately. For example, it implies that if $M\in {\cal T}$ and $N\in {\cal T}$, then $M.N \in {\cal T}$. Along with Constraint~\ref{welldefinedconstraint}, it implies that if $M^A \in {\cal T}^A$, then $M^A\otimes I^B \in {\cal T}^{AB}$. Finally, Assumption~\ref{alltransformations}, along with Assumption~\ref{directproductassumption} and Constraint~\ref{reducedstateconstraint}, implies that if a procedure consists in introducing an ancilla to system $A$, performing some joint transformation on $A$ and ancilla and then throwing away the ancilla, then the corresponding transformation on $A$ alone is $\in {\cal T}^A$. 
 
The fact that transformations have to be well-defined yields one of the main insights of this work. There is a rich interplay between the set of allowed states, the allowed dynamics, and the information processing possibilities that a theory offers. For example, if a theory is modified by enlarging the set of allowed states (adding super-correlated states to quantum theory, perhaps), one might naively think that this must increase the information processing possibilities. However, enlarging the set of allowed states may well have the effect of decreasing the set of allowed transformations, in which case the effect may well be the opposite.

\section{A brief note on ambiguities}\label{ambiguities}

There are a couple of points that deserve a mention here in case it be thought that they cause problems (this section may perhaps be omitted on a first reading). First, two theories may be identical in their structure, that is the sets ${\cal S}$, ${\cal T}$, ${\cal O}$, ${\cal R}$ and ${\cal M}$ of allowed states, transformations, operations, outcomes and measurements, could be mathematically identical in each theory, yet the theories be different physically because the mathematical objects are assigned to different physical objects. For example, a particular preparation device could be associated with one state in one theory and another state in the other theory. 

Second, one theory could be made to look different, that is have different sets ${\cal S}$, ${\cal T}$, ${\cal O}$, ${\cal R}$ and ${\cal M}$, simply because different measurement devices are chosen to correspond to fiducial measurements. Thus in quantum theory the state of a qubit could be specified by the probabilities for the outcomes of spin measurements in the $x$, $y$ and $z$ directions. The set ${\cal S}$ is then a sphere. Equally, the quantum state could be specified by the probabilities for measurements in the $x$, $y$ and $n$ directions, where $\vec{n}=1/\sqrt{2}(\vec{x}+\vec{z})$. In this case, the set ${\cal S}$ is an non-spherical ellipsoid. A fiducial set may even have different numbers of measurements and outcomes. For example, any quantum state can be expressed by giving the probabilities of the outcomes for a single, informationally complete POV measurement \cite{fuchs}. The important thing is that the outcomes of the fiducial measurements in the new formulation are represented by linearly independent vectors in the old formulation. Thus there is an invertible matrix $N$ such that the two formulations are related by $\vec{P}'=N.\vec{P}$, $\vec{R}^{\prime T}=\vec{R}^T . N^{-1}$, and $M'=N.M.N^{-1}$. The theory makes the same predictions since $\vec{R}'.\vec{P}'=\vec{R}.\vec{P}$, and so on.  

The first of these points means that in order to compare the predictions of two theories, one has to know which physical devices different preparations and operations correspond to. But being primarily interested in the information processing properties of theories, we can ignore this issue and concentrate on the structure of the theories. The second point ensures that we can do this unambiguously. The structure of a theory and the conclusions drawn for information processing do not depend on which measurements are chosen for the fiducial set.

\section{Some different theories}\label{sec:some-diff-theor}

It is useful to see examples of theories that can be described in this framework. The most important are classical theories and quantum theory. Two others are GLT and GNST. All of these theories satisfy Assumption~\ref{alltransformations}, which means that each is completely determined by the set ${\cal S}$ of allowed states for each type of system.

\subsection{Classical theories}\label{sec:class-prob-theory}

Suppose that for some particular type of system, the fiducial set can be chosen as a single measurement with $d$ outcomes, and that any (possibly sub-normalized) probability distribution over these outcomes corresponds to a (possibly sub-normalized) allowed state. In this case, the system is \emph{classical}. A classical theory is one for which all systems are classical. The most comprehensive classical theory is the one for which there is a type of system for every $d\geq 1$. For a classical system, ${\cal S}$ is a simplex. Pure states are represented by vectors $\vec{e}_i$, with a 1 for the $i$th component and $0$s elsewhere. The state of a bipartite system of two classical systems is also represented by a vector from a probability simplex, the entries being the joint probabilities for outcomes $i$ and $j$ when the fiducial measurement is performed on each system. An allowed transformation $M$ must map a pure state $\vec{e}_i$ to another allowed state. It is easy to show that each entry of $M$ must be positive, and the sum of each column must be $\geq 0$ and $\leq 1$. In the case that $M$ preserves normalization, it is a stochastic matrix.\footnote{In this work, a stochastic matrix is a not necessarily square matrix, with positive entries, whose columns each sum to $1$.} The set ${\cal R}$ is a hypercube. 

Consider, for example, an ordinary die which can exist in six different deterministic states. The $\vec{P}$ vector is six dimensional and gives the probabilities that the die's uppermost face is $1,2,\ldots,6$. An example of a measurement is one that asks, is the uppermost face $1$ or $2$? The yes outcome corresponds to the vector $\vec{R}=(1,1,0,0,0,0)$. The state of two dice, $A$ and $B$, can be written as a $36$-dimensional vector, whose entries are the probabilities for the uppermost faces being $11,12,\ldots,66$. 

Suppose that the reduced states of the two dice are given by $\vec{P}^A=\vec{P}^B=1/6(1,1,1,1,1,1)$. One possible joint state compatible with $\vec{P}^A$ and $\vec{P}^B$ is a direct product
\begin{align*}
\vec{P}^{AB}&=\vec{P}^A\otimes \vec{P}^B\\
&=\frac{1}{36}\left(\begin{array}{c}1\\1\\\vdots\\1\end{array}\right).
\end{align*}
This corresponds to the two dice being uncorrelated. But another possible joint state with the same reduced states is
\begin{equation*}
\vec{P}^{AB}_{ij}=\left\{\begin{array}{cl}1/6&\ \ i=j,\\0&\ \ \mathrm{otherwise.}\end{array}\right.
\end{equation*}
This corresponds to perfect correlation and obviously cannot be written as a direct product. Of course there is no entanglement or nonlocality in this theory.\footnote{There is nothing difficult in the preceding remarks. But part of the aim of Section~\ref{sec:multi-part-syst} is to deflate the significance of the tensor product rule for combining systems in quantum theory. Thus it is useful to note that a similar rule arises quite naturally in what is essentially classical probability theory. The quantum tensor product rule does not have to be regarded, as it frequently is, as a mysterious replacement for the Cartesian product used in combining deterministic classical states. If quantum states (even pure ones) are more analogous to probabilistic classical states than anything else - in other words if some version of the \emph{epistemic interpretation} of the quantum state is correct - then a tensor product rule is exactly what one would expect. Thus one way of viewing the tensor product is as evidence for the epistemic interpretation.}

\subsection{Quantum Theory (in finite dimensions)}\label{sec:quantum-mechanics-in}

Quantum theory only allows certain types of system. For example,
there are no systems that can be described with two fiducial
measurements each with two outcomes. A qubit can be described by three
fiducial measurements with two outcomes, e.g., spin measurements in
the $x$, $y$ and $z$ directions. Once a set of fiducial measurements
is chosen quantum theory tells us what the allowed states $\vec{P}$
are. In the simple case of a qubit, the set of normalized states is the Bloch sphere. In the case of higher dimensional quantum systems it does not appear to be so easily characterized (except via the usual quantum formalism of course). The transformations that are well-defined, in the sense of Definition~\ref{welldefined}, correspond precisely to the linear completely positive maps. It is usually assumed that any such map corresponds to a physically possible operation, thus Assumption~\ref{alltransformations} is satisfied. Any set of $\vec{R}_i$ with $0\leq \vec{R}_i.\vec{P}\leq 1 \ \forall i \ \forall \vec{P}\in {\cal S}$ and $\sum_i \vec{R}_i.\vec{P}=1\ \forall \vec{P}\in {\cal S}$ is a positive operator-valued measurement in the usual formalism. 

There is nothing new in the fact that quantum states can be represented as real vectors and transformations as matrices acting on these vectors. It is well known that Hermitian operators in $d$ dimensions form a $d^2$-dimensional real vector space, with an inner product given by $\mathrm{Tr}(AB)$. Linear completely positive maps correspond to $d^2\times d^2$ matrices acting on this space. But the present framework does not correspond exactly to this representation 
(e.g., it is possible that $\vec{P}.\vec{P}> 1$), so it is useful to see an example. A qubit whose state is spin up in the $z$-direction can be written
\begin{equation*}
\vec{P}=\left(\begin{array}{c}P(\uparrow|x)\\P(\downarrow|x)\\\hline P(\uparrow|y)\\P(\downarrow|y)\\\hline P(\uparrow|z)\\P(\downarrow|z)\end{array}\right)=\left(\begin{array}{c}1/2\\1/2\\\hline 1/2\\1/2\\ \hline 1\\0\end{array}\right),
\end{equation*}
where $P(\uparrow|x)$ is the probability of obtaining spin up when measuring in the $x$ direction, and so on. It can now be verified that if, for example, spin is measured in the $n$-direction, where $\vec{n}=1/\sqrt{2}(\vec{x}+\vec{z})$, then the up outcome corresponds to the vector
\begin{equation*}
\vec{R}=\left(\frac{1}{2\sqrt{2}},\frac{-1}{2\sqrt{2}}\left|\frac12,\frac12\right|\frac{1}{2\sqrt{2}},\frac{-1}{2\sqrt{2}}\right).  
\end{equation*}
This vector is not unique. Any vector $\vec{R}'=\vec{R}+\vec{C}$, where $\vec{C}.\vec{P}=0\ \forall \vec{P}\in {\cal S}$, represents the same measurement outcome. The unitary transformation usually written as the Pauli matrix $\sigma_z$ would correspond to
\begin{equation*}
M = \left(\begin{array}{cc|cc|cc} 0&1&0&0&0&0\\1&0&0&0&0&0\\ \hline 0&0&0&1&0&0\\0&0&1&0&0&0\\ \hline 0&0&0&0&1&0\\0&0&0&0&0&1\end{array}\right).
\end{equation*}

\subsection{Generalized Non-Signalling Theory}\label{sec:gener-non-sign}

Suppose that for any pair $n,k>1$, there is a corresponding type of single system, whose state can be described by a set of $n$ fiducial measurements, each with $k$ outcomes. Call this an $(n,k)$ system.\footnote{A more general theory would include further types of system with different numbers of outcomes for different fiducial measurements. I ignore this possibility. I do not believe that it would change much beyond introducing uninteresting complications into some of the proofs.} For a single system, allow any state $\vec{P}$, provided the entries of $\vec{P}$ are between $0$ and $1$ and Eq.~\eqref{relativenorm} is satisfied. For multi-partite systems, allow any state $\vec{P}$, provided entries are between $0$ and $1$, Eq.~\eqref{relativenorm} is satisfied, and the no-signalling conditions of Eqs.~\eqref{nosignalling1},\eqref{nosignalling2} are satisfied for all bipartite splittings. The resulting theory is Generalized Non-Signalling Theory. 

\begin{figure}
\scalebox{0.67}{\includegraphics{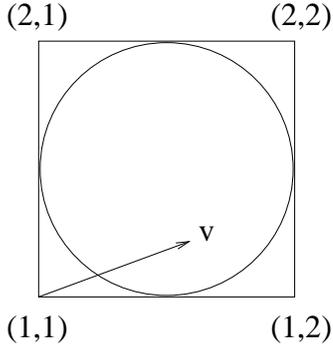}}
\caption[]{The space of normalized states for a gbit in GNST corresponds to the
  square. If the measurements $X=1$ and $X=2$ are associated with spin measurements in the $z$ and $x$ directions, then the space of states for a quantum mechanical qubit corresponds to the circle.\label{squarecircle}}
\end{figure}

It is useful to see some examples of systems in this theory. The
simplest kind of single system has two binary fiducial
measurements. This type of system plays a role somewhat analogous to
that of a classical bit or a qubit, so from hereon it is called a \emph{gbit} (for generalized bit). The space of possible normalized states is
shown in Fig.~\ref{squarecircle}. There are four pure states, which
correspond to the four ways of assigning definite outcomes to the
$X=1$ and $X=2$ fiducial measurements. In the figure, these are
represented by $(1,1)$, $(1,2)$, $(2,1)$, and $(2,2)$, where $(1,2)$,
for example, is the state which returns $a=1$ for the $X=1$
measurement and $a=2$ for the $X=2$ measurement, and is also
represented by $\vec{P}=(1,0|0,1)$. Thus pure states of single
systems have a definite outcome for each fiducial measurement - there
is no uncertainty principle.
As noted in the figure, if the measurements $X=1$ and $X=2$ are
associated with spin measurements in the $z$ and $x$ directions, then
we can include possible states of a qubit in the diagram, and these
form a circle inscribed in the square. Qubits of course have an
extra degree of freedom, namely spin in the $y$ direction. For
$(3,2)$ systems the space of states is a cube, with an inscribed
sphere (the Bloch sphere) representing quantum states.

\begin{figure}
\scalebox{0.50}{\includegraphics{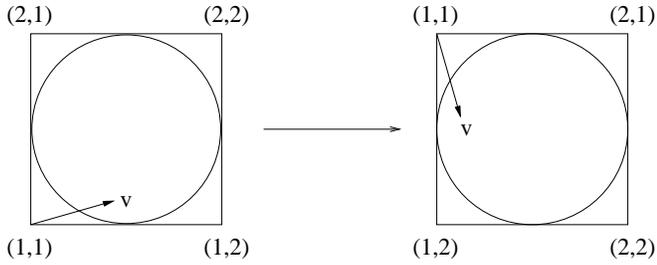}}
\caption[]{An allowed transformation.\label{allowedrot}}
\end{figure}

\begin{figure}
\scalebox{0.50}{\includegraphics{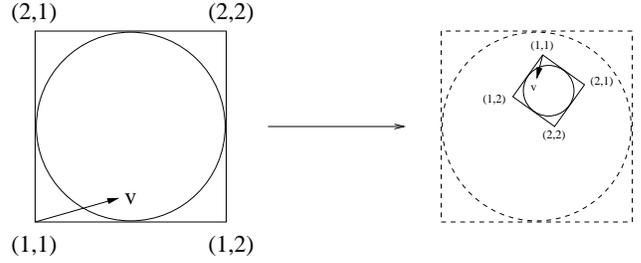}}
\caption[]{Another allowed transformation.\label{allowedshrink}}
\end{figure}

\begin{figure}
\scalebox{0.38}{\includegraphics{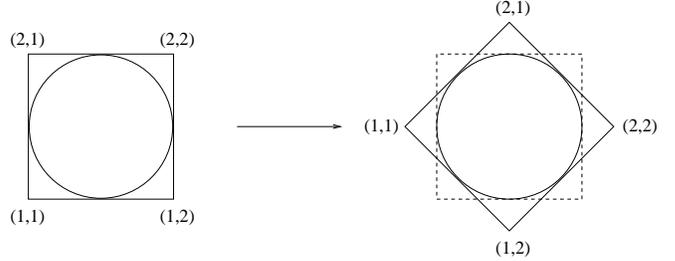}}
\caption[]{A forbidden transformation.\label{notallowed}}
\end{figure}

Consider the possible transformations of a gbit (for simplicity,
restrict attention to those that preserve normalization). An allowed
transformation will transform the square in such a manner that all
points remain in the square, otherwise the transformation is not
well-defined in the sense of Definition~\ref{welldefined}. The
transformations of Figs.~\ref{allowedrot} and \ref{allowedshrink} are allowed.
But the transformation of Fig.~\ref{notallowed} is not allowed. 
Transformations in quantum theory are less restricted because the
requirement is only that points in the circle are transformed into
points in the circle. So a rotation of $\pi/4$, as in
Fig.~\ref{notallowed}, is fine, and indeed corresponds to the well
known $\pi/8$ gate.

It begins to look as if the dynamics of single systems in GNST is
rather simple. Indeed, this is the case. Section~\ref{dynamics}
contains a theorem that states that for single systems in GNST,
allowed transformations correspond essentially to relabellings of
measurements and outcomes, and probabilistic combinations
thereof. Thus in a sense, the dynamics is classical. Despite this, the
dynamics does contain possibilities that quantum dynamics does
not. Consider a $(3,2)$ system, whose space of normalized states is a
cube, with the quantum Bloch sphere inscribed. A possible transformation is a reflection in the center of the sphere. This corresponds to the so called Universal NOT gate of quantum theory, which is not an allowed transformation since it is not completely positive.    

The multi-partite states of GNST are noteworthy in that they
include states that are more nonlocal than quantum theory
allows. For example, given a bipartite system of two gbits, the
following is a possible state.
\begin{align}\label{prdef}
&\begin{split} XY = \left.\begin{array}{c} 11 \\ 12 \\ 21 \end{array}\right\} \rightarrow  P(&a=1,b=1|XY)=\\
&P(a=2,b=2|XY)=1/2,\end{split} \\
&\begin{split} XY = 22  \rightarrow P(&a=1,b=2|XY)=\\
&P(a=2,b=1|XY)=1/2.\end{split}
\end{align}
The correlations obtained from fiducial measurements on this state
return a value of $4$ for the left hand side of the following
inequality
\begin{equation}\begin{split}
P(a=b|11) &+ P(a=b|12) \\
&+ P(a=b|21) + P(a\neq b|22) \leq 3
\end{split}\end{equation}
(this is the CHSH inequality \cite{chsh} written in a slightly
different form than usual). These correlations cannot be obtained from
measurements on any quantum state since by Tsirelson's theorem
\cite{tsirelson}, quantum states can only reach a maximum of
$2+\sqrt{2}$.\footnote{These non-signalling super-quantum correlations
  were written down by Khalfi and Tsirelson \cite{kt}, and were
  independently introduced by Popescu and Rohrlich \cite{pr}. Other
  examples of super-quantum correlations, involving more measurements
  or parties, are given in Ref.~\cite{barrettetal}. The latter are
  also allowed in GNST.} 

Information processing in GNST is discussed in
Section~\ref{informationprocessing}. The theory's permissiveness with
respect to states implies that some things can be achieved that are
impossible in quantum theory. These include 1-2 oblivious transfer,
van Dam's scheme for the easy solution of communication complexity
problems \cite{vandam}, and a kind of super-quantum memory. The
restricted nature of the dynamics, however, implies that there is no
teleportation or super-dense coding. The theorems of Section~\ref{dynamics}
give evidence that computation is no better than classical.

\subsection{Generalized Local Theory}\label{sec:gener-local-mech}

Suppose that, as in GNST, for any pair $n,k>1$, there is a
corresponding type of system, whose state can be defined with $n$
fiducial measurements with $k$ outcomes. As in GNST, all $\vec{P}$ with entries between $0$ and $1$ satisfying Eq.~\eqref{relativenorm} are allowed states. The only multi-partite states allowed, however, are
those for which the fiducial measurements return local
(non-Bell-violating) correlations. This defines Generalized Local Theory.

As in GNST, the pure states of single systems in this theory are
those that have a deterministic outcome for each fiducial
measurement. Since multi-partite states are local with respect to
fiducial measurements, the pure states of a multi-partite system are
precisely those in which each subsystem is in a deterministic pure
state. An arbitrary state of a multi-partite system is a convex
mixture of these. It follows that no state in this theory can violate
a Bell inequality, even if non-fiducial measurements are
performed. Hence the name.  

GLT is more general than quantum theory in allowing arbitrary
single system states, but more restricted in not allowing nonlocal
states. As described in Section~\ref{informationprocessing}, GLT
allows 1-2 oblivious transfer. Computation in GLT, however, is
efficiently simulable by a classical computer.  

\subsection{Other possibilities}\label{sec:other-possibilities}

There are other possibilities that would be interesting to
investigate. For example,
\begin{enumerate}
\item A theory that is essentially quantum theory but with only
  separable states allowed.
\item A theory in which the state of a single system must be a quantum
  state, but in which the state of a multi-partite system can be
  anything, as long as the no-signalling principle and the restriction
  that the reduced states for the individual subsystems must be
  quantum are satisfied. The latter idea has been investigated in
  Ref.~\cite{barnumetal}, where it is shown, amongst other things,
  that Tsirelson's theorem still holds.
\end{enumerate}

\section{Generic properties of theories}\label{sec:some-basic-prop}

One of the reasons for introducing a framework encompassing many different theories is that it is interesting to identify properties of theories that are generic, in the sense that they are shared by all or most theories in the framework. Some features, usually thought of as specifically quantum, are present in all theories in our framework except theories that are classical (in the sense of Section~\ref{sec:class-prob-theory}). Thus classical theories are very special! These features include the fact that mixed states do not always have a unique decomposition into pure states, and a no-go theorem for universal cloning. More exact statements of these claims are given in this section. Proofs are in Appendix~\ref{genericfeatures}. It is tedious to write always \emph{all theories in the framework}, so from hereon this is shortened to \emph{all theories}, taking the assumptions of Section~\ref{sec:fram-prob-theor} as read.

First,
\begin{theorem}\label{mixedstatedecompositions}
Suppose that for a particular type of system, every mixed state has a unique decomposition into pure states and $\vec{0}$. Then the system is classical.
\end{theorem}

The next theorem concerns the disturbance of systems on measurement and is due in part to Howard Barnum and Alex Wilce \cite{howardalex}. Say that a transformation \emph{disturbs} a state $\vec{P}$ if there is no constant $c$ such that $M.\vec{P} = c\,\vec{P}$. This means that, conditioning on the outcome corresponding to this transformation, the state is no longer $\vec{P}$. A transformation is non-disturbing if no pure state is disturbed and an operation $\{M_i\}$ is non-disturbing if all $M_i$ are non-disturbing.
\begin{theorem}\label{measurementsdisturb}
For any system, let $V$ be the vector space in which states are defined, and let $V_S$ be the subspace spanned by ${\cal S}$. Then $V_S$ can be written as a direct sum, $V_S=\bigoplus_i V_i$, where the $V_i$ are subspaces of $V_S$, such that
\begin{enumerate}
\item Every pure state $\vec{P}$ is contained in some $V_i$.
\item A non-disturbing transformation is of the form $M=\bigoplus_i e_i I_i$, where $0\leq e_i\leq 1$, and $I_i$ is the identity on $V_i$.
\end{enumerate}
\end{theorem}
It follows that non-disturbing operations have the same outcome probabilities for pure states in the same $V_i$, and thus cannot distinguish them. It is easy to show that a system is classical if and only if each $V_i$ contains exactly one pure state. For a quantum system without superselection rules, $V_S$ cannot be further decomposed into a direct sum. Non-disturbing operations have the same outcome probabilities for all pure states, each transformation being proportional to the identity on $V_S$. An example of such an operation would be to toss a coin, without interacting with the system at all, and to output the result. For a quantum system with superselection rules, pure states from the same sector are elements of the same $V_i$, and different sectors correspond to different $V_i$.

Theorem~\ref{measurementsdisturb} has implications for cloning. Cloning refers to the following procedure:  
\begin{enumerate}
\item Begin with a system $A$ in a pure state. Denote its state $\vec{P}$.
\item Introduce a system $B$ of the same type, prepared in a standard state $\vec{Q}$. The state of the joint system is $\vec{P}\otimes \vec{Q}$.
\item A joint transformation $M$ acts on the pair of systems such that the final state is $(M)(\vec{P}\otimes \vec{Q})\propto \vec{P}\otimes \vec{P}$.  
\end{enumerate}
A \emph{deterministic universal cloning procedure} always succeeds and works on all pure states. It implies the existence of a normalization-preserving $M$ and a state $\vec{Q}$ such that $(M)(\vec{P}\otimes \vec{Q}) = \vec{P}\otimes \vec{P}$ for all pure $\vec{P}$. A \emph{probabilistic universal cloning procedure} is allowed to output a fail outcome, but conditioned on success, the final state must be $\vec{P}\otimes \vec{P}$. There must be a non-zero probability of success for all pure states $\vec{P}$. This type of cloning implies the existence of a non-zero $M$ such that $(M)(\vec{P}\otimes \vec{Q})=c\vec{P}\otimes \vec{P}$ for all pure $\vec{P}$, where $c$ can vary with $\vec{P}$, and $0 < c \leq 1$.

\begin{theorem}\label{nocloning}
With the exception of classical systems, there is no probabilistic universal cloning procedure.
\end{theorem}
This of course implies that with the exception of classical systems, there is no deterministic universal cloning procedure. 

Theorems~\ref{mixedstatedecompositions}, \ref{measurementsdisturb} and \ref{nocloning} apply even to classical theories if extended to mixed states. Thus there are mixed states with a non-unique decomposition into mixed states. All transformations disturb at least one mixed state unless they are proportional to the identity.\footnote{This is not at all surprising if put into more prosaic terms. Consider that a die is in a state such that the probability of each face being uppermost is $1/6$. Suppose that the die is measured, to find out which face is uppermost, and the value $1$ found. Then, if it is assumed that the measurement operation was done in the most obvious way, the state after measurement is not $1/6(1,1,1,1,1,1)$, but $(1,0,0,0,0,0)$. Of course the measurement operation may be such that the die is recast after the outcome is obtained, resulting in a final state of $1/6(1,1,1,1,1,1)$. But then an initial state of $(1,0,0,0,0,0)$ would be disturbed.} And cloning of classical mixed states is impossible.\footnote{Suppose that Alice prepares a die in one of two ways, each corresponding to a probability distribution over the different faces. The first prepares, say, the state $1/12(6,2,1,1,1,1)$ and the second the state $1/6(1,1,1,1,1,1)$. The die is given to Bob who is required to perform a cloning operation. This means that Bob must prepare another die such that if Alice used the first preparation, its state is $1/12(6,2,1,1,1,1)$, and if she used the second, then $1/6(1,1,1,1,1,1)$. Furthermore, if the dice are measured after Bob's operation, the results \emph{must not be correlated}. This last clause prevents Bob from using a device that simply reads the uppermost face of the die and prepares another in the same state. It is easy to see that even if Bob's cloning procedure is allowed to be probabilistic, he cannot do it.} One possible interpretation of these remarks is as further evidence that quantum pure states are more akin to classical mixed states than classical pure states. 

There are many other questions concerning properties that are common to all theories, or all except classical theories. In Ref.~\cite{newbroadcast}, the quantum no-broadcasting theorem is generalized to arbitrary non-classical theories within a framework closely related to the present one. It can also be shown that all theories in the framework have an infinite de Finetti theorem \cite{barrettleifer}, and that polynomially-sized computations in these theories can be simulated classically in polynomial space \cite{barrettpspace}. Features such as these can be regarded as arising solely from the assumptions that were made in setting up the framework. 

\section{Dynamics in GNST and GLT}\label{dynamics}

Part of the motivation of this work is to consider which features of a theory, in particular those features related to information processing, arise from which assumptions. It is particularly interesting if significant features, such as the no-cloning theorem above, arise from very minimal assumptions and are thus shared by a broad class of theories. Another part of the motivation is to investigate theories that are different from those we already know about. These theories need not even be empirically adequate; a compare and contrast exercise will still be useful to learn more about those theories that \emph{are} empirically adequate. Thus the next two sections are devoted to a detailed investigation of GNST and GLT.  

In Section~\ref{sec:gener-non-sign} the dynamics of a gbit was briefly discussed. There are four pure states of a gbit, corresponding to the four ways of assigning definite outcomes to the two measurements. The space of normalized states is a square, with a normalization-preserving transformation being a linear transformation of this square. Let us consider more general types of system in GNST and GLT, but continue to focus on normalized systems and normalization-preserving transformations, i.e., operations corresponding to a single matrix, $\{M\}$. For this section and the next, \emph{transformation} means \emph{normalization-preserving transformation}, with the investigation of probabilistic transformations left for future work. 

\begin{figure}
\scalebox{0.67}{\includegraphics{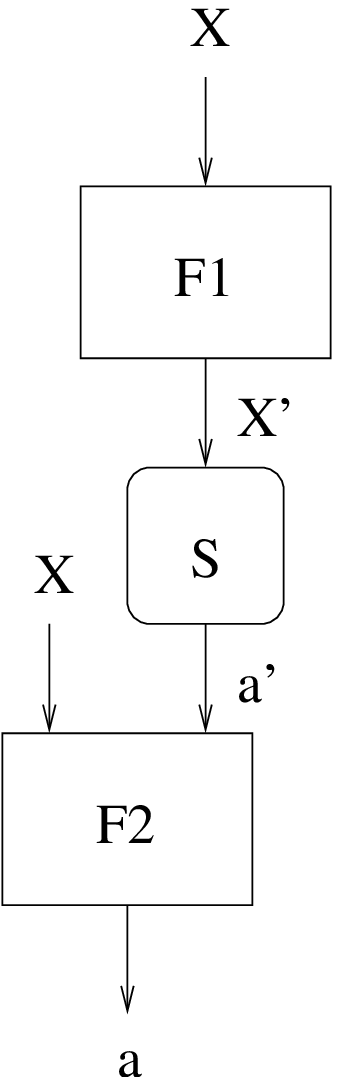}}
\caption[]{Transformations of single systems in GNST and GLT can always be represented as the appending of classical circuits as shown here, or as convex combinations of transformations of this type. If a fiducial measurement $X$ is performed on the transformed system, this can be thought of as performing fiducial measurement $X'$ on the original system, where $X'=F1(X)$ for some function $F1$. When measurement $X'$ is performed on the original system, outcome $a'$ is obtained with some probability. The probability of obtaining outcome $a$ for the measurement $X$ on the transformed system is equal to the probability of obtaining an outcome $a'$ such that $a=F2(X,a')$, for some function $F2$.\label{singlesystemcircuits}}
\end{figure}

\begin{figure}
\scalebox{0.67}{\includegraphics{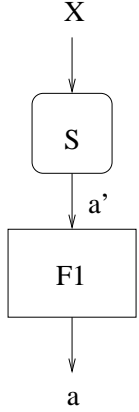}}
\caption[]{Measurements on single systems in GNST or GLT can always be performed via a procedure of the type illustrated here, or via a convex combination of such procedures. First, fiducial measurement $X$ is performed and outcome $a'$ is obtained. The outcome $a$ of the complete measurement is then $a=F1(a')$. This result applies to measurements with an arbitrary number of outcomes.\label{singlesystemmeasurements}}
\end{figure}

The space of normalized states of an $(n,k)$ system is a polytope, the vertices corresponding to pure states. Pure states are of the form 
\begin{equation*}
\vec{P}= (0\ldots 1\ldots 0 | 0\ldots 1 \ldots 0 | \ldots ).
\end{equation*}
Allowed transformations must take points in the polytope to points in the polytope. This condition is so restrictive that the following theorem holds.
\begin{theorem}\label{singlesystemtheorem}
Normalization-preserving transformations of single systems in GNST or GLT, thought of as active, correspond to passive transformations that simply relabel fiducial measurements and outcomes, or to convex combinations of such. Equivalently, for a transformation of an $(n,k)$ system, the matrix $M$ representing the transformation can be written
\[
M=\left( \begin{array}{c|c|c} 
M_{11} & \cdots & M_{1n} \\ \hline
\vdots & \ & \vdots \\ \hline
M_{n1} & \cdots & M_{nn} \end{array}\right), 
\]
where $M_{ij}$ is a $k\times k$ matrix, and where $M_{ij} = \alpha_{ij}S_{ij}$, for $S_{ij}$ a stochastic matrix, $0\leq \alpha_{ij} \leq 1$, and $\sum_j \alpha_{ij}=1$.
\end{theorem}
A useful pictorial representation of this theorem is given in Fig.~\ref{singlesystemcircuits}.
A related result is
\begin{theorem}\label{singlesystemcorollary}
The only measurements on single systems in GNST or GLT are fiducial measurements, possibly with outcomes relabelled, or correspond to convex combinations of such. 
\end{theorem}
Theorem~\ref{singlesystemcorollary} is illustrated pictorially in Fig.~\ref{singlesystemmeasurements}.
The proofs of Theorems~\ref{singlesystemtheorem} and \ref{singlesystemcorollary} are contained in Appendix~\ref{trivialdynamicsproof}.

\begin{figure}
\scalebox{0.67}{\includegraphics{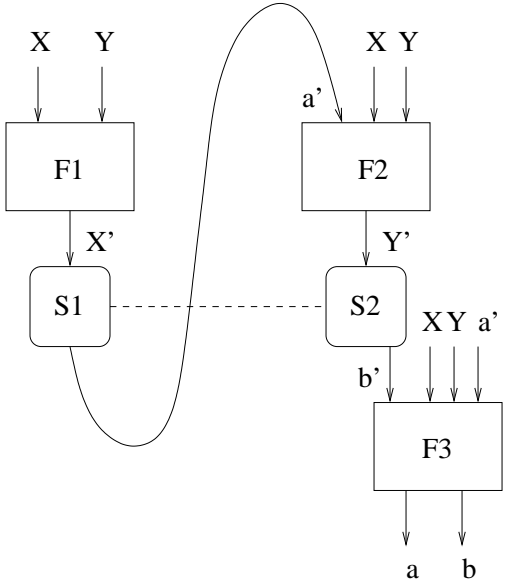}}
\caption[]{In GNST, transformations on bipartite systems comprised of two gbits can always be represented by the appending of classical circuits as shown here, or by a similar construction inverted with respect to the two systems, or by a convex combination of such. For the construction shown here, this means that if fiducial measurements $X,Y$ are performed on the transformed system, one may think of this as first performing a fiducial measurement $X'$ on one half of the original system, where $X'=F1(X,Y)$. This gives an outcome $a'$. Then, perform a fiducial measurement $Y'$ on the other subsystem, where $Y'=F2(X,Y,a')$. This gives an outcome $b'$. The final outcome pair $(a,b)$ is determined by a function $F3$ of $X$, $Y$, $a'$ and $b'$.\label{bipartitecircuits}}
\end{figure}

\begin{figure}
\scalebox{0.67}{\includegraphics{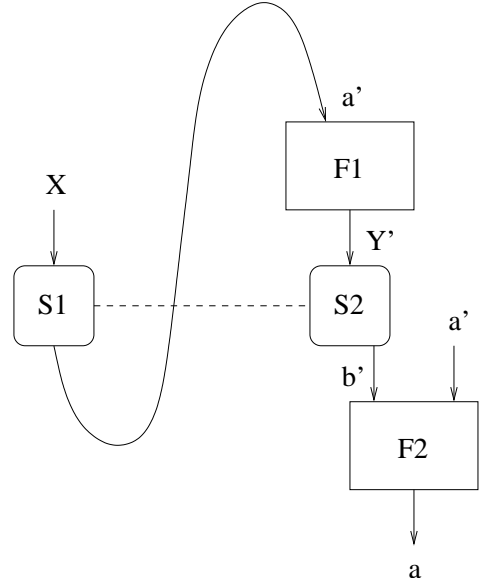}}
\caption[]{In GNST, measurements on bipartite systems of two gbits can always be carried out by a procedure like that illustrated here, by a similar procedure inverted with respect to the two subsystems, or by a convex combination of such. For the procedure shown here, this means that first, a fiducial measurement $X$ is performed on one subsystem, and outcome $a'$ is obtained. Then fiducial measurement $Y'$ is performed on the other subsystem, where $Y'=F1(a')$, and outcome $b'$ is obtained. The outcome $a$ of the complete measurement is given by $a=F2(a',b')$. This result applies to measurements with an arbitrary number of outcomes.\label{bipartitemeasurements}}
\end{figure}

In the case of GNST, the following theorem holds for a bipartite system of two gbits, and suffices to characterize the normalization-preserving transformations of such a system. 
\begin{theorem}\label{bipartitetheorem}
Consider a system of two gbits in GNST, and suppose that a normalization-preserving transformation is performed. Suppose that this transformation is followed by the fiducial measurements $X,Y$ on the two subsystems, with outcomes $a,b$. The joint probability of obtaining outcomes $a,b$ is equal to that obtained from a convex combination of procedures of the following kind. First, perform a fiducial measurement $X'$ on one of the gbits, where $X'$ may depend on $X$ and $Y$. Denote the outcome $a'$. Then perform a fiducial measurement $Y'$ on the other gbit, where $Y'$ may depend on $X,Y$ and on $a'$. Denote the outcome $b'$. The final outcome pair $(a,b)$ is a function of $X$, $Y$, $a'$ and $b'$.
\end{theorem}
Of course this theorem can also be expressed in terms of a formal constraint on the transformation matrix $M$, but in this case it is more complicated and less enlightening. Theorem~\ref{bipartitetheorem} may also be understood pictorially, as in Fig.~\ref{bipartitecircuits}.
\begin{theorem}\label{bipartitecorollary}
In GNST, the only measurements on bipartite systems comprised of two gbits correspond to convex combinations of procedures of the following kind. First, perform a fiducial measurement $X$ on one of the gbits, obtaining an outcome $a'$. Then perform a fiducial measurement $Y$ on the other gbit, where $Y$ may be a function of $a'$, obtaining an outcome $b'$. The final outcome is a function of $a'$ and $b'$.
\end{theorem}
Theorem~\ref{bipartitecorollary} is illustrated in Fig.~\ref{bipartitemeasurements}.
The proof of Theorem~\ref{bipartitetheorem} is given in Appendix~\ref{trivialdynamicsproof}. It is an open question whether similar theorems hold for transformations and measurements on arbitrary multi-partite systems in GNST. It can be shown that in GLT, there definitely do exist possibilities for measurements and transformations on multi-partite systems that do not reduce to one of the forms presented in this section.

The proofs in Appendix~\ref{trivialdynamicsproof} also make clear the following. The most fine-grained measurements on single systems in GNST or GLT can be represented by a set of vectors $\vec{R}_i$, such that each $\vec{R}_i$ has one element between 0 and 1 and the rest 0. Such an $\vec{R}_i$ is analogous to an effect in quantum theory that is proportional to a 1-dimensional projector. A set of $\vec{R}_i$ is analogous to a non-degenerate projective measurement if each $\vec{R}_i$ is a basis vector (one element 1 and the rest 0) and $\sum_i \vec{R}_i.\vec{P} = 1 \ \forall \vec{P}\in {\cal S}$. The corresponding measurement is simply a fiducial measurement, with an $\vec{R}_i$ for each outcome. It is then immediate that, at least with respect to these measurements, there is no Kochen-Specker theorem for single systems in GNST or GLT. Not only is it possible to assign definite outcomes to these measurements in a non-contextual fashion, but each such assignment is in fact an allowed state of the theory. Nonetheless, both GNST and GLT exhibit a different kind of contextuality, introduced by Spekkens \cite{spekkenscontextuality} and termed \emph{preparation contextuality}. Readers are referred to Ref.~\cite{spekkenscontextuality} for discussion of preparation contextuality. Given the definition, the proofs for GNST and GLT are obvious.

\section{Information processing}\label{informationprocessing}

Using the results obtained for dynamics in GNST and GLT, the information processing possibilities of each theory can be investigated. Rather than attempt something like a general theory of information, this section contains remarks concerning some obvious tasks. Note that there has already been some work investigating the information processing properties of PR boxes, considered merely as abstract correlations. van Dam has shown that they are very powerful for communication complexity problems \cite{vandam}, and this result has recently been extended to noisy PR boxes in Ref.~\cite{brassardetal}. Others have claimed to show how to do oblivious transfer \cite{boxot} and bit commitment \cite{boxcommit} using PR boxes. However, as pointed out in Ref.~\cite{short}, the fact that these latter works consider PR boxes only as abstract correlations means that they make assumptions that may not hold in any theory that allows PR boxes.\footnote{One such assumption that as far as I know has not been pointed out is that the shared boxes are trusted to behave like PR boxes by both parties. But one may reasonably ask where did they come from? By whom were they distributed?} In general, a theory with well defined dynamics is needed before cryptography, or indeed other types of information processing, such as computation, can be discussed. GNST is such a theory.

The first results concern teleportation and super-dense coding (the quantum versions of these tasks were introduced in Refs.~\cite{teleportation} and \cite{densecoding}). The natural analogue of a quantum mechanical singlet in the GNST is a state which, when fiducial measurements are performed, produces the PR box correlations:
\begin{align*}
&\begin{split} XY = \left.\begin{array}{c} 11 \\ 12 \\ 21 \end{array}\right\} \rightarrow  P(&a=1,b=1|XY)=\\
&P(a=2,b=2|XY)=1/2,\end{split} \\
&\begin{split} XY = 22  \rightarrow P(&a=1,b=2|XY)=\\
&P(a=2,b=1|XY)=1/2.\end{split}
\end{align*}
It can be shown that these correlations represent a pure state - that is a vertex of the polytope of states for two gbits. Further, all vertices of this polytope are either local deterministic correlations (product pure states) or are equivalent to the PR box under local transformations \cite{barrettetal}. 
\begin{theorem}
It is impossible to teleport an unknown gbit using a single shared PR box.
\end{theorem}
\emph{Proof}. This follows easily from Theorem~\ref{bipartitecorollary}. In order to teleport an unknown gbit, Alice must perform some operation or sequence of operations on the gbit and her half of the shared PR box. Without loss of generality, whatever she does may be represented as a single joint measurement, with $m$ outcomes, on the two subsystems. But this measurement can be represented as a convex combination of procedures like that of Fig.~\ref{bipartitemeasurements}. Such a procedure will always begin, either by measuring $X=1$ or $X=2$ on the gbit, or by measuring the PR box. In the former case, no information is gained about the value for the other measurement on the gbit and teleportation cannot possibly succeed on all pure states. In the latter case, the shared PR box collapses into a product state which cannot achieve teleportation. \hfill$\square$

\begin{theorem}
A single shared PR box cannot be used for super-dense coding.
\end{theorem}
\emph{Proof}. This follows from Theorems~\ref{singlesystemtheorem} and \ref{bipartitecorollary}. Super-dense coding would require that there are four different operations that Alice can perform on her gbit such that, when it is sent to Bob, he can determine unambiguously which was performed by a joint measurement on the two gbits now in his possession. It is easy to see that this is not possible. \hfill$\square$  

\subsection{Cryptography}

\begin{theorem}
In GNST, key distribution is possible.
\end{theorem}
\emph{Proof}. Key distribution can be achieved in GNST using an Ekert-style protocol \cite{ekert}, in which Alice and Bob first share $n$ pairs of gbits, with each pair in the PR box state. They then test some of their shared systems, to make sure that they really are PR box states, i.e., that they have not been disturbed en route by an eavesdropper. Finally, they measure each remaining gbit pair, using the fiducial measurements $X=1$ and $Y=1$. Assuming that they share perfect PR box states, their measurement outcomes will be perfectly correlated and can be used as a secret key. This protocol is secure because PR box states have a property of being monogamous, much as the entanglement of a singlet is monogamous in quantum theory. Thus consider a tripartite system shared between Alice, Bob and Eve. If Alice's and Bob's reduced state is the PR box state $\vec{P}^{AB}_{PR}$, then the global state must be of the form $\vec{P}^{AB}_{PR}\otimes \vec{P}^{E}$. The outcome of any measurement performed by Eve is uncorrelated with Alice's and Bob's outcomes. The fact that the PR box correlations are monogamous was shown in Ref.~\cite{barrettetal}.\hfill$\square$

Recall Theorem~\ref{measurementsdisturb}, which implies that except for classical systems, there are pure states (lying in the same subspace $V_i$), which cannot be distinguished by non-disturbing operations. This motivates
\begin{conjecture}
In any non-classical theory, secure key distribution is possible.
\end{conjecture} 

Finally,
\begin{theorem}
1-2 oblivious transfer can be implemented securely in both GNST and GLT.
\end{theorem}
\emph{Proof}. In a 1-2 oblivious transfer (introduced in Ref.~\cite{obtrans}), Alice must submit 2 bits to Bob in such a manner that Bob can choose to learn either one of the bits or the other, but not both. There is also a security requirement against Alice, who must not be able to learn which of the bits Bob chose. That this task is impossible to implement securely in quantum theory is shown in Ref.~\cite{lo}. To implement this task in GNST or GLT, Alice sends a gbit to Bob, in a pure state, with the two bits encoded in the outcomes for the $X=1$ measurement and the $X=2$ measurement. Theorem~\ref{singlesystemcorollary} ensures that any strategy employed by Bob is equivalent to his measuring either $X=1$ or $X=2$, or to measuring $X=1$ with some probability $p$ and $X=2$ with probability $1-p$. Thus the protocol is secure against Bob. That it is secure against Alice follows from the fact that, by the no-signalling principle, she cannot determine which measurement Bob did. \hfill$\square$

In classical cryptography, it is known that 1-2 oblivious transfer is equivalent to oblivious transfer \cite{brasscreprob}, and that either can be used to implement arbitrary secure distributed computation \cite{killian}. In particular, either can be used to implement bit commitment, hence coin tossing. However, one cannot assume that the standard reductions of classical cryptography hold in a different theory such as GLT, GNST or quantum theory. Thus it is open whether other two-party cryptographic tasks, such as oblivious transfer, bit commitment or coin tossing, can be implemented securely in GNST or GLT. 

\subsection{Computation}

For any of the theories in the framework, a natural model of computation may be defined, based on the classical and quantum circuit models. I introduce this model only informally. A particular circuit is assumed to act on $n$ systems, each of the same type, initially prepared in a product state corresponding to the problem input. Instead of $k$-bit or $k$-qubit gates, there are transformations that act jointly on $k$ systems. At the end of the computation, the fiducial measurement $X=1$ is performed on each system in order to obtain the output. For a particular theory, it may not be the case that bipartite and single system transformations together are universal, as they are in classical and quantum theory. Thus transformations that act jointly on $k$ systems for any $k>2$ are allowed. But for any circuit family $C_n$, there must exist some finite $k$ such that all transformations act on $k$ systems or fewer. In addition, it may not be the case that any particular type of system (such as a gbit) is universal for computation in a given theory. So one should keep in mind that circuits may act on other types of system. Finally, in order to define a notion of polynomial time, say, the usual caveats must be assumed. For example, it should be possible for a classical Turing machine to output a description of the $i$th circuit in time polynomial in $i$.

\begin{theorem}  
In GLT, any computation can be simulated efficiently by a probabilistic classical computer.
\end{theorem}
Proof. In GLT, any allowed state of $n$ systems can be written as a convex combination of local deterministic, or pure product, states, in which each system has a definite outcome for each fiducial measurement. A classical simulation of the GLT computation works by storing, at any given time, a local deterministic state of the $n$ systems. This requires an amount of memory linear in $n$, rather than the exponential amount needed to store a complete description of an arbitrary convex combination. An allowed transformation $T$, acting on $k$ systems, must take local deterministic states of the $k$ systems to other allowed states of GLT, which in turn are convex combinations of local deterministic states:
\[
T(\vec{P}^{LD})=\sum_i p_i \vec{P}^{LD}_i,
\]
where superscript $LD$ indicates a local deterministic state. The classical computer simulating the GLT computation simply updates the stored state $\vec{P}^{LD}$ to $\vec{P}^{LD}_i$ with probability $p_i$. When the final $X=1$ measurements are performed, the stored local deterministic state will determine the classical computer's output.\hfill$\square$   

The computational power of GNST is at present unclear. But it is known that it is very powerful for communication complexity problems.
\begin{theorem}
In GNST, bipartite communication complexity problems require only constant communication, provided the parties share sufficient PR boxes.
\end{theorem}
Recall that in a bipartite communication complexity scenario, two separated parties each receive an input, and their task is to compute some joint function of their inputs. Their goal is to minimize the amount of communication. van Dam has shown that if the two parties have a supply of shared PR boxes, then any communication complexity problem can be solved with only constant communication \cite{vandam}. This result has recently been strengthened: it continues to hold even if the shared PR boxes are noisy, provided the amount of noise is not too great \cite{brassardetal}. Contrast the situation in quantum theory, where the inner product problem is known to require $n$ bits of communication to be solved exactly, even with unlimited shared singlets \cite{innerproduct}.

Finally,
\begin{theorem}
Super-quantum memory. In GNST, it is possible to store a $2^n$-bit string in only $n$ gbits. Although the whole string cannot be recovered, it is possible to recover the $i$th bit without error.
\end{theorem}
\emph{Proof}. Suppose that the $i$th bit of the $2^n$-bit string we wish to store is given by $f(i_1,\ldots,i_n) \in \{0,1\}$, where $i_1\ldots i_n$ is the binary representation of $i$. Let $X_1,\ldots,X_n \in \{0,1\}$ be fiducial measurements on the $n$ gbits and $a_1,\ldots,a_n \in\{0,1\}$ the outcomes. (It is easier for this proof to let $X_j$ and $a_j$ take values in $\{0,1\}$ instead of in $\{1,2\}$ as elsewhere.) To store the string, prepare a state of $n$ gbits such that 
\begin{equation}\label{memorystate}\begin{split}
P(a_1,&\ldots,a_n|X_1,\ldots,X_n)=\\
&\left\{\begin{array}{cc} 1/2^{n-1} & \qquad a_1 \oplus \cdots \oplus a_n = f(X_1,\ldots,X_n) \\
0 & \qquad\mathrm{otherwise}\end{array}\right. ,\end{split}
\end{equation}
where $\oplus$ represents addition mod 2. In order to recover the $i$th bit of the stored string, simply perform the measurement $X_j=i_j$ on each gbit and sum the outcomes mod 2. One may check that the state of Eq.~(\ref{memorystate}) is an allowed state, since it is normalized and non-signalling. Note that it is indeed impossible to store a $2^n$ bit string in only $n$ qubits such that any bit may be recovered. Bounds on quantum memory are derived in Ref.~\cite{ambainisetal}.\hfill$\square$

\section{Discussion}\label{discussion}

\subsection{The framework}

The framework introduced allows investigation of theories different from either quantum or classical theories. The general idea is that quantum theory can be better understood by viewing it in a context of different possibilities. More specific motivations include: 
\begin{enumerate}
\item to understand the links between general physical principles and information processing;
\item to stimulate the study of computation in models that are more general than quantum theory;
\item to address Popescu's and Rohrlich's question of why quantum theory does not allow the PR box correlations;
\item to shed light on the interpretive problems of quantum theory by viewing those in a more general context;
\item to stimulate research into axioms for quantum theory.  
\end{enumerate}

As regards single systems the framework is very general indeed. It should be emphasized in particular that linearity of transformations is not assumed, but is derived from the fact that the vector $\vec{P}$ is by definition a complete description of the system.\footnote{So what of nonlinear modifications of quantum mechanics? These modifications are nonlinear in the sense that they involve a nonlinear Schr\"{o}dinger equation. In this case, the usual density matrix is no longer a complete description of a quantum system, since the evolution of a system will in general depend not only on the density matrix, but on the particular decomposition into pure states (assuming a proper mixture). If the description of the state is expanded until it is complete, then the action of the dynamics on this new expanded state description will be linear. But such a theory will in general violate one or more of the other assumptions. A list of references on nonlinear quantum theories is given in Ref.~\cite{svetlichnynonlinear}, and computation in this context is considered in Ref.~\cite{abrams98}.} The most important requirements are that local operations commute (Assumption~\ref{localopscommute}), and the global state assumption (Assumption~\ref{globalstateassumption}), both involving the manner in which separate systems combine to make joint systems. These imply a tensor product rule. 

One of the interesting things to emerge from the framework is that certain features, usually thought of as specifically quantum, are possessed by all theories except classical theories. These include the non-unique decomposition of mixed states into pure states, the existence of sets of pure states that cannot be distinguished with non-disturbing operations, and the impossibility of even probabilistic universal cloning. Thus rather than regard quantum theory as special for having these features, a better attitude may be to regard classical theories as special for not having them. 

How reasonable are Assumptions \ref{localopscommute} and \ref{globalstateassumption}? Commutativity of local operations is arguably part of what it means to talk about separate systems. In a theory where it fails, any measurement or transformation is essentially a measurement or transformation on all systems at once. It is no longer obvious how to define a reasonable model of computation - how should resources be counted? The case for assuming the commutativity of local operations is also strengthened by the fact that in a spacetime framework, it can be independently motivated by special relativity. It is slightly more difficult to regard the global state assumption as independently compelling. Thus an interesting direction in which to extend this work would be to generalize the framework further by dropping this assumption.

\subsection{The tensor product rule}

It is interesting to compare the derivation of the tensor product rule with that of Fuchs \cite{fuchs}. Without going into too much detail, Fuchs assumes that local measurements on two separate systems, $A$ and $B$, are represented by positive operator-valued measures on Hilbert spaces $H_A$ and $H_B$. He derives a Gleason-like theorem \cite{gleason, busch, fuchsgleason} which states that the joint state of the two systems can be represented by an operator on the tensor product Hilbert space $H_A\otimes H_B$, with joint probabilities for outcomes of local measurements given by the standard trace rule.

As Fuchs acknowledges, the proof does not establish that the operator describing the joint state has to be positive, but only that it has to be positive with respect to local measurements. A consistent theory that is not ruled out would allow the state to be negative with respect to some joint measurements (the Bell basis measurement, for example), but would not allow such measurements. Furthermore, the assumption that local operations commute and the global state assumption are both implicit in Fuchs' analysis. Without the latter, the possibility remains that there are extra degrees of freedom, not accessible via local measurements, that are not described by an operator on the tensor product Hilbert space.

It follows that Fuchs' conclusion is not stronger than the tensor product rule derived in this paper. The latter may be regarded as a generalization of Fuchs' proof to the case in which the subsystems $A$ and $B$ are not necessarily quantum. 

\subsection{Information theory, GNST and GLT}

In addition to describing general properties of the framework, I investigated in detail two particular theories, GNST and GLT. I focussed on the information processing possibilities in these theories. One of the most interesting things to have emerged is that there is a trade-off between the states of a theory and the allowed dynamics. This arises for the simple reason that an allowed transformation must take allowed states into allowed states. Thus the dynamics of both GNST and GLT is very simple for single systems. In GNST, a similar result holds for the simplest kind of bipartite system. The surprising consequence is that GNST is less powerful than quantum theory in many ways, despite including super-quantum correlations. For example, teleportation and super-dense coding are impossible. It is already clear that computation in GLT can be simulated efficiently classically, while the computational power of GNST remains open. Another open question is whether secure bit commitment is possible in either theory. Despite these remarks, it is surprising how many features of quantum theory have analogues in GNST. These obviously include the generic features demonstrated in Section~\ref{genericfeatures}, along with entanglement and nonlocality. But they also include things I have not discussed in detail, such as the distinction between sharp and unsharp measurements, and preparation contextuality. (Other authors have also found features of quantum theory reproduced in other contexts. Masanes et al. \cite{masanes} show that various features, including a no-cloning theorem, are present in all theories that are nonlocal and non-signalling. Spekkens has introduced a toy theory that contains a remarkably wide range of quantum phenomena \cite{spekkens}, although note that this theory is not contained in our framework as it does not allow arbitrary convex combinations of states.)

As mentioned above, one of the motivations of this work is to stimulate the study of computation in models that are more general than quantum theory. Some authors have already considered computation in non-standard theories. However, these theories are often modifications of quantum theory that appear to have both unphysical consequences and immense computational power. It is suspected that quantum theory with a nonlinear Schr\"{o}dinger equation is very powerful, enabling the solution of NP-complete problems in polynomial time, for example.\footnote{In Ref.~\cite{abrams98}, it is claimed that nonlinear quantum theory can solve NP-complete and even \#\! P-complete problems efficiently. Aaronson complains \cite{aaronson} that in this particular case it is difficult to evaluate whether exponential precision is required.} Aaronson has considered various modifications of quantum theory, including a model that assumes the ability to postselect measurement outcomes, and a hidden variable model in which the history of hidden states can be read out by the observer \cite{aaronson1,aaronson3}. Various authors have considered classical and quantum computation in the presence of closed timelike curves \cite{brun,bacon}. Most recently, Aaronson and Watrous have shown that BQP with closed timelike curves is equivalent to PSPACE \cite{aaronsoninprep}. The framework introduced in this paper is the natural place to investigate computation in theories that are different from quantum theory, yet not obviously physically unreasonable or immensely powerful. I suggest that NP-complete problems cannot be solved efficiently by any theory in the framework. I also raise the following
\begin{conjecture}
A quantum computer can simulate computation in any other theory in the framework with at most polynomial overhead.
\end{conjecture}
The intuition behind this is that quantum theory achieves in some sense an optimal balance of allowed states and dynamics.

\subsection{Interpretation}

On the face of it, many theories that can be written down in the present framework have similar interpretive issues as quantum theory, if one tries to understand them in a way that goes beyond the purely operational. Consider a universe in which some theory other than quantum or classical (GNST perhaps) is verified in laboratory experiments. The denizens of such a universe would be having debates in many ways similar to the debates that surround quantum theory. Is a pure state better understood as a complete description of individual reality, as representing an ensemble, or as representing the degrees of belief of some agent? 

Suppose that the inhabitants of this universe attempt to extend the theory to include a description of the measuring apparatus, and of the interaction between system and apparatus. This is always possible in classical and quantum theory. In quantum theory, this fact is expressed in the idea that the \emph{Heisenberg cut} can be moved upwards indefinitely. Are classical and quantum theories special in this regard, or can this be done in any theory? 

Even when the inhabitants succeed in constructing a measurement theory along these lines, it is plausible that many theories will have a \emph{measurement problem}. In these theories, the system and apparatus are typically in some entangled state after interaction. Some inhabitants may suggest hidden variables or some kind of collapse dynamics. Does any theory admit an Everettian interpretation, or is there a special feature of quantum theory that is necessary for this to work?

I won't discuss these issues any further. I have raised them hoping that considering interpretive issues in a framework more general than quantum theory might give a new lease of life to the quantum debates.

\subsection{Axioms}

Aside from Hardy's derivation \cite{hardy}, what different ways are there of uniquely identifying quantum theory from the other theories in the framework by adding as few extra assumptions as possible? Several have pushed the idea that a quantum state is best understood as a summary of an agent's degrees of belief about the outcomes of future measurements on a system \cite{cavesfuchs,cfs,cfscertainty}. From this standpoint, Fuchs has argued that the formalism of quantum theory should be understood as a constraint on these degrees of belief, hopefully to be derived via a small number of postulates, along with an argument that any rational agent must accept \cite{fuchs}. Spekkens has also argued for an epistemic constraint as a foundational principle for quantum theory, although for Spekkens, beliefs are about underlying ontic states of a system rather than future measurement outcomes \cite{spekkens}. 

Clifton, Bub and Halvorson (CBH) have taken a different approach and derived at least part of quantum theory from the assumption of (i) a no-signalling principle, (ii) a no-broadcasting principle, and (iii) the impossibility of secure bit commitment \cite{cliftonbubhalvorson}.\footnote{Whether they succeed in deriving the full structure of quantum theory is debatable. But they do establish the existence of non-commuting measurements and of entanglement.} CBH assume a $C^*$-algebraic framework, which is broad enough to include classical theories and quantum theory, but is not as broad as the framework presented here. An open question is whether something like CBH's proof would go through in the broader framework, or whether there is some theory (GNST perhaps) that satisfies (i)-(iii) and is clearly not quantum.

\acknowledgments

Research at Perimeter Institute is supported in part by the Government of Canada through NSERC and by the Province of Ontario through MEDT. I should very much like to thank Howard Barnum, Lucien Hardy, Nick Jones, Matthew Leifer, Lluis Masanes, Stefano Pironio, Sandu Popescu, Valerio Scarani, Tony Short, Rob Spekkens and Alex Wilce for useful discussions.

\emph{Note added.} Related independent work has appeared recently in Ref.~\cite{entswapping}.

\appendix

\section{Proof of linearity of transformations}\label{linearityproof}

The proof in this appendix is adapted from that of Hardy in Ref.~\cite{hardy}. It is included to keep this work self-contained. 

A transformation is a map from allowed states of a system to allowed states. The map satisfies Eq.(\ref{preservecombinations}), reproduced here:
\begin{equation}\label{preservecombinationsappendix}\begin{split}
f\left(\sum_i q_i \vec{P}_i\right) &= \sum_i q_i f(\vec{P}_i) \\
&\forall P_i\in{\cal S},\ \mathrm{for}\ 0\leq q_i\leq 1,\ \sum_i q_i=1.
\end{split}\end{equation}
The map should also satisfy
\[
f(\vec{0}) = \vec{0}.
\]
(This follows from the interpretation of unnormalized states. Recall that if a particular outcome $i$ of some operation occurs with probability $q<1$, then we associate with that outcome an unnormalized vector $\vec{P}$. Each entry of $\vec{P}$ gives the joint probability of obtaining outcome $i$ for the original operation, and outcome $j$ for a fiducial measurement performed immediately afterwards. Thus if $q=0$, it follows that the associated $\vec{P}=\vec{0}$. By definition, an entry in the vector $f(\vec{0})$ represents the joint probability of getting the following outcomes in sequence: outcome $i$ for the original operation, then whatever outcome it is that corresponds to the transformation $f$, and then outcome $j$ for a fiducial measurement. But these probabilities must all be zero if the probability of outcome $i$ is zero.)  


Writing the first of the above equations with $i=1,2$, and setting $\vec{P}_2=\vec{0}$, gives
\[
f(q\vec{P}) = q f(\vec{P}) \qquad \forall \vec{P}\in{\cal S},\ \mathrm{for}\ 0\leq q\leq 1.
\]
Suppose that $\vec{P}$ is a pure state $\in {\cal S}$. Pure states are by definition normalized. If $r>1$, then $f(r\vec{P})$ is initially undefined because $r\vec{P}\notin{\cal S}$, so we are free to stipulate that 
\[
f(r\vec{P}) = r f(\vec{P})\qquad \forall\vec{P}\in{\cal S},r\geq 0.
\]
Define ${\cal S}_+$ as the set of all vectors that can be written in the form $r\vec{P}$ with $\vec{P}\in {\cal S}$ and $r\geq 0$. It is a convex cone \cite{maths}. Eq.~(\ref{preservecombinationsappendix}) can be extended slightly:
\begin{equation}\label{preservecombinationsextended}
f\left(\sum_i r_i \vec{P}_i\right) = \sum_i r_i f(\vec{P}_i)
\qquad \forall P_i\in{\cal S}_+\ \mathrm{for}\ r_i\geq 0.
\end{equation}
Now suppose that
\begin{equation}\label{arbitrarylinear}
\vec{P}=\sum_i s_i \vec{P}_i, 
\end{equation}
where $\vec{P},\vec{P}_i \in {\cal S}_{+}$, and the $s_i$ are real. Let $i\in A_{-}$ if $s_i<0$ and $i\in A_{+}$ if $s_i\geq 0$. Rewrite Eq.~(\ref{arbitrarylinear}) as
\[
\vec{P} + \sum_{i\in A_{-}} |s_i| \vec{P}_i = \sum_{i\in A_{+}} s_i \vec{P}_i.
\]
Each side is a conic combination of vectors in ${\cal S}_{+}$, thus Eq.~(\ref{preservecombinationsextended}) applies, and rearranging we get
\[
f(\vec{P}) = \sum_i s_i f(\vec{P_i}).
\]
Finally, for any vector $\vec{Q}\notin {\cal S}_{+}$, $f(\vec{Q})$ can be defined uniquely by linear extension if $\vec{Q}$ lies in the subspace spanned by ${\cal S}$. The action of $f$ on the rest of the vector space is arbitrary but may be defined to be linear.\hfill$\square$

\section{Derivation of tensor product rule}\label{tensorproductproof} 

As discussed in the main text, the state of a joint system $AB$ can be written
\[
\vec{P}^{AB} \equiv \left(\begin{array}{c} P(a=1,b=1|X=1,Y=1)\\P(a=1,b=2|X=1,Y=1)\\ \vdots \\ \hline P(a=1,b=1|X=1,Y=2) \\ P(a=1,b=2|X=1,Y=2) \\ \vdots \\ \hline \vdots \end{array}\right).
\]

\emph{Proof of Theorem~\ref{tensorproducttrivialtheorem}}. This theorem is trivial. Let $\vec{P}^{AB}\in V^{AB}$, $\vec{P}^A\in V^A$ and $\vec{P}^B\in V^B$. Define the vector $\vec{Q}^{AB}_{ijkl}$ as the vector with a $1$ for the entry corresponding to the joint outcome $ij$ of the joint fiducial measurement $kl$, and $0$s elsewhere. Similarly $\vec{Q}^{A}_{ik}$ and $\vec{Q}^{B}_{jl}$. Now identify $\vec{Q}^{AB}_{ijkl}$ with $\vec{Q}^A_{ik}\otimes \vec{Q}^B_{jl}$ and extend linearly.\hfill$\square$  

\emph{Proof of Theorem~\ref{directproducttheorem}}. Consider a joint system $AB$. For each of the fiducial measurements that define the state of system $B$, there must be at least one operation on the joint system $AB$ that corresponds to performing that measurement. Let this operation for the $j$th fiducial measurement be characterized by the set of matrices $\{M_{ij}\}$, where there is a value of $i$ for each outcome and $j$ is fixed. When the transformation $M_{ij}$ acts on $AB$, the resulting state is the unnormalized state $\vec{P}^{AB}_{ij}\in{\cal S}^{AB}$. The corresponding reduced state for $A$ is the unnormalized state $\vec{P}^{A}_{ij}$. By Constraint~\ref{reducedstateconstraint}, $\vec{P}^{A}_{ij}\in{\cal S}^A$. If a fiducial measurement is now performed on $A$, the state $\vec{P}^{A}_{ij}$ gives the (unnormalized) probabilities for the different outcomes. It follows that $\vec{P}^{AB}$ can be written in the form
\begin{equation}\label{pabrep}
\vec{P}^{AB}= \sum_{ij} \vec{P}^A_{ij}\otimes \vec{Q}^B_{ij},
\end{equation}
with $\vec{P}^A_{ij}\in{\cal S}^A$ and $\vec{Q}^B_{ij}$ as above.
Now consider a vector $\vec{U}\otimes \vec{W}\in V^{AB}$, with $\vec{W}\in{\cal S}^B$ but $\vec{U}\perp{\cal S}^A$, where this means that $\vec{U}$ is orthogonal to all vectors in ${\cal S}^A$. From Eq.(\ref{pabrep}) it follows that $(\vec{U}\otimes \vec{W}).\vec{P}^{AB}=0$. A similar result holds if $\vec{W}\perp{\cal S}^B$ and $\vec{U}\in{\cal S}^A$. Thus $\vec{P}^{AB}$ lies in the subspace of $V^{AB}$ that is spanned by vectors from ${\cal S}^A\otimes {\cal S}^B$. Eq.~\eqref{directproduct} follows. The vectors on the right hand side of this equation can be assumed normalized, since any multiplying factor can be subsumed into the corresponding $r_i$. They can be assumed pure, since a mixed state can always be expressed as a convex combination of pure states and $\vec{0}$. But any term with $\vec{0}$ will not contribute. Theorem~\ref{directproducttheorem} follows.\hfill$\square$

\emph{Proof of Theorem~\ref{tensorprodfortransforms}}. Consider a joint system $AB$ and a transformation $T^A$ of system $A$ alone. $T^A$ corresponds to a matrix $M^A$ such that $\vec{P}^A\rightarrow \vec{P}^{\prime A} = M^A . \vec{P}^A$. The aim is to determine the effect of this transformation on the joint state $\vec{P}^{AB}$. From Section~\ref{sec:stat-transf}, this will correspond to a matrix $\tilde{M}^A$ such that $\vec{P}^{AB}\rightarrow \vec{P}^{\prime AB} = \tilde{M}^A.\vec{P}^{AB}$. But what is the relation between $M^A$ and $\tilde{M}^A$?

Consider the following procedure. First, the transformation $T^A$ is applied. Then fiducial measurements are performed on systems $A$ and $B$. The (unnormalized) joint probabilities for the outcomes of these measurements are then the entries of the vector $\vec{P}^{\prime AB}$. However, by Assumption~\ref{localopscommute}, the ordering of operations on systems $A$ and $B$ does not matter. Thus the following procedure is equivalent. First, a fiducial measurement is performed on system $B$. Note that the reduced state of system $A$ conditioned on a particular outcome for this measurement is defined by the vector $\vec{P}^{AB}$. Next, the transformation $T^A$ is performed on system $A$. Finally, a fiducial measurement is performed on system $A$. 

In the second procedure, we know how to apply the transformation $T^A$, since it is enough to consider its action on system $A$ alone, and we know that $\vec{P}^A\rightarrow \vec{P}^{\prime A} = M^A . \vec{P}^A$. We obtain 
\begin{align*}
\vec{P}^{\prime AB}_{ijkl} &= \sum_{i'k'} (M^A)_{ik;i'k'} \vec{P}^{AB}_{i'jk'l}\\
&=\sum_{i'k'j'l'} (M^A)_{ik;i'k'} \delta_{jj'}\delta_{ll'} \vec{P}^{AB}_{i'j'k'l'}.
\end{align*}
But
\begin{equation*}
(M^A\otimes I^B)_{ijkl;i'j'k'l'} = (M^A)_{ik;i'k'} \delta_{jj'}\delta_{ll'},
\end{equation*}  
thus
\begin{equation*}
\vec{P}^{\prime AB} = (M^A\otimes I) \vec{P}^{AB}.
\end{equation*}
This holds for all $\vec{P}^{AB}\in {\cal S}^{AB}$, and the action of $T^A$ on vectors $\vec{P}^{AB}\notin {\cal S}^{AB}$ is arbitrary. It follows that we lose no generality in identifying
\begin{equation*}
\tilde{M}^A = M^A \otimes I^B.
\end{equation*}\hfill$\square$

\section{Generic Features}\label{genericfeatures}

This appendix contains proofs of the results of Section~\ref{sec:some-basic-prop}. 

\emph{Proof of Theorem~\ref{mixedstatedecompositions}}. Consider a particular type of system in some theory. Suppose that the subspace spanned by allowed states of the system has dimension $d$ and that every mixed state has a unique decomposition into pure states and $\vec{0}$. The only convex set with this property is a simplex with $d+1$ vertices. One of these vertices is the state $\vec{0}$. It is always possible to find an invertible linear transformation $N$ such that the other vertices are transformed into the vectors $(1,0,0,\ldots,0)$, $(0,1,0,\ldots,0)$, and so on. Recall from Section~\ref{ambiguities} that if this transformation acts on the set ${\cal S}$, then the theory is not changed, since $\vec{R}^T\rightarrow\vec{R}^T.N^{-1}$ and $M\rightarrow N.M.N^{-1}$ for measurements and transformations. Hence the system is classical. \hfill$\square$

\emph{Proof of Theorem~\ref{measurementsdisturb}}. Consider a system with a set of allowed states ${\cal S}$, spanning $V_S$, and let $d$ be the dimension of $V_S$. Choose a set of $d$ distinct pure states $\{\vec{P}_1,\ldots,\vec{P}_d\}$ that are linearly independent and collectively span $V_S$. Suppose that a particular transformation is non-disturbing. Its action on each of the $\vec{P}_i$ is given by $M.\vec{P}_i = c_i \vec{P}_i$ with $0\leq c_i \leq 1$. If the system is classical, then ${\cal S}$ is a simplex and the set $\{\vec{P}_1,\ldots,\vec{P}_d\}$ must contain all the pure states. Since the $\vec{P}_i$ are linearly independent, the $c_i$ can be chosen independently without contradiction. For any other type of system, there are at least $d+1$ pure states. Consider a pure state $\vec{Q}$ that is not contained in the set $\{\vec{P}_1,\ldots,\vec{P}_d\}$. If the transformation is non-disturbing, then $M.\vec{Q}=e\vec{Q}$ with $0\leq e\leq 1$. Since $\{\vec{P}_1,\ldots,\vec{P}_d\}$ is a basis for $V_S$, $\vec{Q}$ has a unique decomposition of the form $\vec{Q}=\sum_i d_i \vec{P}_i$, where at least two of the $d_i$ are non-zero. If $d_j$ and $d_k$ are non-zero, then $c_j=c_k=e$. Thus $M$ acts as $e$ times the identity on the subspace of $V_S$ spanned by $\vec{P}_j$ and $\vec{P}_k$. By repeating this reasoning for every pure state $\vec{Q}$, the set $\{\vec{P}_1,\ldots,\vec{P}_d\}$ can be divided into subsets such that (i) if $\vec{P}_j$ and $\vec{P}_k$ are in the same subset, then $c_j=c_k$ for any non-disturbing transformation, and (ii) if $\vec{P}_j$ and $\vec{P}_k$ are in different subsets then there is no pure state $\vec{Q}$ such that both $d_j$ and $d_k$ are non-zero. Each subset defines a subspace $V_i$ of $V_S$ and the theorem follows.\hfill$\square$

\emph{Proof of Theorem~\ref{nocloning}}. Theorem~\ref{nocloning} is proven using Theorem~\ref{measurementsdisturb}. We show that if there is a probabilistic universal cloning procedure, then for any two pure states $\vec{P}_1$ and $\vec{P}_2$, there is a non-disturbing transformation $M'$ such that $|M'.\vec{P}_1|\neq |M'.\vec{P}_2|$. This in turn implies that the system is classical.

Suppose that there is a standard state $\vec{Q}$ and a transformation $M$ such that for each pure state $\vec{P}$, $M(\vec{P}\otimes\vec{Q})= c\vec{P}\otimes\vec{P}$. The number $c$ may vary with $\vec{P}$ but is $>0$ for all $\vec{P}$. Consider a procedure in which a system is in the state $\vec{P}_1$ or $\vec{P}_2$, an ancilla is added in the standard state $\vec{Q}$, and the cloning operation $\{M,F\}$ performed on the joint system. The transformation $M$ corresponds to the success outcome and $F$ to the fail outcome. If $\vec{P}_1$ and $\vec{P}_2$ are different states there must be some operation $\{ N_1,N_2 \}$ such that $|N_1.\vec{P}_1|\neq |N_1.\vec{P}_2|$. If cloning succeeded, perform this operation on the ancilla. Output the result and throw away the ancilla. 

This entire procedure may be regarded as an operation on the system alone (see the remarks following Assumption~\ref{alltransformations}). It can be written $O'=\{ M'_1, M'_2, F' \}$, where $M'_1$ corresponds to successful cloning followed by the $N_1$ outcome, $M'_2$ corresponds to successful cloning followed by the $N_2$ outcome, and $F'$ corresponds to failed cloning. By construction, each of $M'_1$ and $M'_2$ is non-disturbing and $|M'_i.\vec{P}_1|\neq |M'_i.\vec{P}_2|$ for at least one of $i=1,2$.  

Recalling Theorem~\ref{measurementsdisturb}, it follows that $V_S=\bigoplus_i V_i$, with each $V_i$ containing only one pure state, hence the system is classical. \hfill$\square$

\section{Dynamics in GNST and GLT}\label{trivialdynamicsproof}

This appendix contains proofs of Theorems~\ref{singlesystemtheorem}, \ref{singlesystemcorollary} and \ref{bipartitetheorem}, all of which concern dynamics in GNST or GLT.
 
\emph{Proof of Theorem~\ref{singlesystemtheorem}}. This theorem concerns transformations of single systems in either GNST or GLT. A transformation of an $(n,k)$ system can be written
\begin{widetext}
\begin{equation}
\left(\begin{array}{c} P'(a=1|X=1)\\ \vdots \\ P'(a=k|X=1) \\ \hline \vdots \\ \hline P'(a=1|X=n) \\ \vdots \\ P'(a=k|X=n) \end{array}\right) =
\left(\begin{array}{c|c|c} 
\begin{array}{ccc}\ & \ & \ \\ \ & M_{11} & \ \\ \ & \ & \ \end{array} &
\begin{array}{ccc}\ & \ & \ \\ \ & \cdots & \ \\ \ & \ & \ \end{array} &
\begin{array}{ccc}\ & \ & \ \\ \ & M_{1n} & \ \\ \ & \ & \ \end{array} \\
\hline
\begin{array}{ccc}\ & \ & \ \\ \ & \vdots & \ \\ \ & \ & \ \end{array} &
\begin{array}{ccc}\ & \ & \ \\ \ & \  & \ \\ \ & \ & \ \end{array} &
\begin{array}{ccc}\ & \ & \ \\ \ & \vdots & \ \\ \ & \ & \ \end{array} \\
\hline
\begin{array}{ccc}\ & \ & \ \\ \ & M_{n1} & \ \\ \ & \ & \ \end{array} &
\begin{array}{ccc}\ & \ & \ \\ \ & \cdots & \ \\ \ & \ & \ \end{array} &
\begin{array}{ccc}\ & \ & \ \\ \ & M_{nn} & \ \\ \ & \ & \ \end{array} 
\end{array}\right) 
\left(\begin{array}{c} P(a=1|X=1)\\ \vdots \\ P(a=k|X=1) \\ \hline \vdots \\ \hline P(a=1|X=n) \\ \vdots \\ P(a=k|X=n) \end{array}\right).
\end{equation}
\end{widetext}
The transformation matrix is $M$, an $nk\times nk$ matrix. If the fiducial measurement $X=1$ has $k$ outcomes, then the top $k$ rows of this matrix determine the probabilities of outcomes when the $X=1$ measurement is performed on the transformed state $\vec{P}'$. Denote the $k\times nk$ submatrix consisting of these rows $M_1$. The next $k$ rows are associated with the fiducial measurement $X=2$, so denote the corresponding submatrix by $M_2$, and so on. The first $k$ columns of $M_i$ multiply into those components of $\vec{P}$ that correspond to the fiducial measurement $X=1$ being performed. Denote the $k\times k$ subsubmatrix consisting of these columns $M_{i1}$. Similarly $M_{i2}$, and so on. Note that each row in $M$, considered as a vector $\vec{R}$, must represent a possible yes/no measurement. This is because if the transformation acts on a state $\vec{P}$, then $\vec{R}.\vec{P}$ gives the corresponding entry in the transformed state $\vec{P}'$, which must be between 0 and 1 for all $\vec{P}\in {\cal S}$. Furthermore, when the transformation is normalization-preserving, the rows $\vec{R}_j$ from a particular $M_i$ satisfy $\sum_j \vec{R}_j.\vec{P} = 1$, whenever $\vec{P}$ is normalized. Hence the rows from a particular $M_i$ correspond to a multiple-outcome measurement. One way of performing this measurement is simply to perform the transformation $M$ first, and then to perform fiducial measurement $X=i$.

There is some redundancy in a measurement vector $\vec{R}$, and in the matrix $M$. If $\vec{R}.\vec{P}=\vec{R}'.\vec{P} \ \forall \vec{P}\in {\cal S}$, then $\vec{R}$ and $\vec{R}'$ represent the same measurement. In particular, if $\vec{R}'=\vec{R}+\vec{C}$, where $\vec{C}.\vec{P}=0 \ \forall \vec{P}\in {\cal S}$, then $\vec{R}$ and $\vec{R}'$ represent the same measurement. An example of such a $\vec{C}$ is
\begin{equation*}
\vec{C} = ( 1,\ldots, 1 | -1,\ldots, -1 | 0, \ldots,0 | \ldots ),
\end{equation*}
where $\vec{C}.\vec{P}=0 \ \forall \vec{P}\in {\cal S}$ is ensured by the normalization of $\vec{P}$. The first step in the proof is to show that any $\vec{R}$ is equivalent in this sense to an $\vec{R}'$ with all components $\geq 0$. 

For this, consider the set of allowed normalized states. This is precisely the set of vectors satisfying the conditions
\begin{align}
\sum_i P(a=i|X=j)&=\sum_{i} P(a=i|X=k) \quad \forall j,k,\label{normconditions}\\
P(a=i|X=j)&\geq 0 \quad \forall i,j,\label{posconditions}\\
\sum_i P(a=i|X=1) &= 1. \label{overallnorm}
\end{align}
Define ${\cal S}_{+}$ as the set of vectors of the form $r\vec{P}$, with $r\geq 0$ and $\vec{P}\in {\cal S}$, and note that in the case of GNST or GLT, ${\cal S}_{+}$ is a polyhedral cone \cite{maths}. It can also be defined as the set of vectors satisfying conditions (\ref{normconditions}) and (\ref{posconditions}). The defining inequalities (\ref{posconditions}) can each be written in the form $\vec{C}_i.\vec{P}\geq 0$, where $\vec{C}_i$ is a constant vector with a 1 in the $i$th position and $0$s elsewhere. The equalities (\ref{normconditions}) can each be written as the conjunction of two inequalities: $\vec{D}_j.\vec{P}\geq 0$ and $\vec{D}_j.\vec{P}\leq 0$ for some constant $D_j$. Define ${\cal R}_{+}$ as the set of vectors $\vec{R}$ such that $\vec{R}.\vec{P}\geq 0 \ \forall \vec{P}\in {\cal S}_{+}$. This is the set of unnormalized measurements and is the dual cone to ${\cal S}_{+}$. It can be shown that if a polyhedral cone is defined by $\{\vec{P}: \vec{A}_i.\vec{P}\geq 0 \ \forall i\}$, then the dual cone is equal to the conic hull of the vectors $\vec{A}_i$. Thus elements of ${\cal R}_{+}$ can be written
\begin{equation}\label{badform}
\vec{R} = \sum_i \lambda_i \vec{C}_i + \sum_j \mu_j \vec{D}_j,
\end{equation}
where $\lambda_i\geq 0$ and $\mu_j$ can be positive or negative. Finally, the vectors $\vec{D_j}$ all satisfy $\vec{D}_j.\vec{P}=0 \ \forall \vec{P}\in {\cal S}_{+}$. Hence any $\vec{R}$ of this form is equivalent to an $\vec{R}$ of the form
\begin{equation}\label{goodform}
\vec{R} = \sum_i \lambda_i \vec{C}_i, 
\end{equation}
and without loss of generality, the components of $\vec{R}$ can be assumed $\geq 0$. This applies both to $\vec{R}$ considered as a measurement and to $\vec{R}$ considered as a row of a transformation matrix $M$.

Assume, then, that $M$ is written in a form with all entries $\geq 0$. To conclude the proof, note that $M$ acting on any properly normalized state (satisfying both Eqs.~(\ref{normconditions}) and Eq.~(\ref{overallnorm})) must result in a state that is also properly normalized. This implies the following. Consider the matrix $M_{ij}$. Denote the sum of the elements in the first column by $S^{ij}_1$, the sum of the elements in the second column by $S^{ij}_2$, and so on. Then $S^{ij}_1=S^{ij}_2=\cdots=S^{ij}_k$ and $\sum_j S^{ij}_1=1$. Hence the matrix $M_{ij}$ is of the form $\alpha_{ij}$ times a stochastic matrix, with $0\leq \alpha_{ij}\leq 1$ and $\sum_j \alpha_{ij}=1$. One may easily check that any transformation that is equivalent to a procedure of the form of Fig.~\ref{singlesystemcircuits} is represented by a matrix of this form with $\alpha_{ik}=1$ for some $k$ and $\alpha_{ij}=0$ for $j\neq k$. Hence we have obtained the general result that any allowed $M$ is a convex combination of transformations of the form of Fig.~\ref{singlesystemcircuits}. \hfill$\square$

\emph{Proof of Theorem~\ref{singlesystemcorollary}}. Let an $m$-outcome measurement on an $(n,k)$ system have outcomes corresponding to $\vec{R}_1,\ldots,\vec{R}_m$, and construct the $m\times nk$ matrix
\[
N=\left( \begin{array}{c} \vec{R}_1^T \\ \vdots \\ \vec{R}_m^T \end{array} \right).
\]
Denote the submatrix consisting of the first $k$ columns of $N$ by $N_1$, that consisting of the next $k$ columns by $N_2$, and so on. The same arguments as in the proof of Theorem~\ref{singlesystemtheorem} can be used to establish that $N$ can be chosen such that all entries are $\geq 0$. Then use the fact that $\sum_i\vec{R}_i.\vec{P}=1$ for normalized $\vec{P}$, and arguments similar to those in the proof of Theorem~\ref{singlesystemtheorem}, to establish that $N_i=\alpha_i S_i$ for $0\leq \alpha_i\leq 1$, $\sum_i\alpha_i =1$, and $S_i$ stochastic. The theorem follows.\hfill$\square$

\emph{Proof of Theorem~\ref{bipartitetheorem}}. Begin as before by showing that without loss of generality, the matrix $M$ can be taken to have all entries $\geq 0$. This part of the proof is identical, except that to conditions (\ref{normconditions}), (\ref{posconditions}) and (\ref{overallnorm}), one should add the no-signalling constraints
\begin{align}
\sum_j &P(a=i,b=j|X=k,Y=1) = \nonumber\\
&\sum_{j} P(a=i,b=j|X=k,Y=2)\quad\forall i,k \label{nosigfortwogbits1}\\
\sum_i &P(a=i,b=j|X=1,Y=l) = \nonumber\\
&\sum_{i} P(a=i,b=j|X=2,Y=l)\quad\forall j,l. \label{nosigfortwogbits2}
\end{align}
Like the conditions (\ref{normconditions}), these constraints can be written as the conjunction $\vec{D}_j.\vec{P}\geq 0$ and $\vec{D}_j.\vec{P}\leq 0$, and $\vec{R}$ can be written in the form of Eq.~(\ref{badform}), hence in the form of Eq.(\ref{goodform}). Now impose that $\vec{P}'=M.\vec{P}$ is normalized for any allowed normalized $\vec{P}$, that is any $\vec{P}$ that satisfies conditions (\ref{normconditions}), (\ref{posconditions}), (\ref{overallnorm}), (\ref{nosigfortwogbits1}), and (\ref{nosigfortwogbits2}). Proving that any such $M$ represents a convex combination of transformations of the form of Fig.~\ref{bipartitecircuits} (or the reversed form with respect to the two subsystems) is a tedious brute force exercise that is omitted. As with Theorem~\ref{singlesystemcorollary}, the proof of Theorem~\ref{bipartitecorollary} is a straightforward variation.\hfill$\square$

\end{document}